%% file: main.tex
\documentclass[onecolumn]{aa} 

\usepackage{amsmath}
\usepackage{graphicx}
\usepackage{physics}
\usepackage{nicefrac}
\usepackage[colorinlistoftodos]{todonotes}
\usepackage[colorlinks=true, allcolors=blue]{hyperref}
\usepackage[normalem]{ulem}

\usepackage{amssymb} 
\usepackage{booktabs}
\usepackage{multirow}
\usepackage{natbib}
\usepackage[nolist]{acronym}

\usepackage{siunitx}
\DeclareSIUnit \parsec {pc}
\DeclareSIUnit \arcsecondfull {arcsec}
\DeclareSIUnit \year{yr}
\DeclareSIUnit \day{day}
\DeclareSIUnit \hour{hr}
\DeclareSIUnit \radiant{rad}
\DeclareSIUnit \degfull{deg}
\DeclareSIUnit \erg {erg}
\DeclareSIUnit \eV {eV}
\DeclareSIUnit \Lsun {L_\odot}
\DeclareSIUnit \Msun {M_\odot}
\DeclareSIUnit \AstroUnit {au}
\DeclareSIUnit \steradian {sr}

\include{data_macros}

\begin{document}

\title{Adding Gamma-ray Polarimetry to the Multi-Messenger Era}
\subtitle{Prospects of joint gravitational--wave and gamma--ray polarimetry studies}

   \author{Merlin Kole
          \inst{1}
          \and
          Francesco Iacovelli
          \inst{2,}\inst{3}
          \and
          Michele Mancarella
          \inst{2,}\inst{3,}\inst{4,}\inst{5}
          \and
          Nicolas Produit
          \inst{6}
          }

   \institute{ Department of Nuclear and Particle Physics, University of Geneva, 24 Quai Ernest-Ansermet, 1205 Geneva, Switzerland\\
              \email{merlin.kole@unige.ch}
         \and
             Département de Physique Théorique and Center for Astroparticle Physics, Université de Genève, 24 quai Ansermet, CH–1211 Genève 4, Switzerland
             \and
             Gravitational Wave Science Center (GWSC), Université de Genève, CH–1211 Genève, Switzerland
             \and
             Dipartimento di Fisica ``G. Occhialini'', Universit\'a degli Studi di Milano-Bicocca, Piazza della Scienza 3, 20126 Milano, Italy
             \and
             INFN, Sezione di Milano-Bicocca, Piazza della Scienza 3, 20126 Milano, Italy
        \and
        Department of Astronomy, University of Geneva, Chemin d'Ecogia, 1290 Versoix, Switzerland
             }

\date{}

\label{firstpage}

  \abstract
   {The last decade has seen the emergence of two new fields within astrophysics: gamma--ray polarimetry and gravitational wave (GW) astronomy. The former, which aims to measure the polarization of gamma--rays in the energy range of 10's to 100's of keV, from astrophysical sources, saw the launch of the first dedicated polarimeters such as GAP and POLAR. Due to both a large scientific interest as well as their potential large signal--to--noise ratios, Gamma--ray bursts (GRBs) are the primary source of interest of the first generation of polarimeters. Polarization measurements are theorized to provide a unique probe of the mechanisms at play in these extreme phenomena. On the other hand, GW astronomy started with the detection of the first black hole mergers by LIGO in 2015, followed by the first multi--messenger detection in 2017.}
   {While the potential of the two individual fields has been discussed in detail in the literature, the potential for joint observations has thus far been ignored. In this article, we aim to define how GW observations can best contribute to gamma--ray polarimetry and study the scientific potential of joint analyses. In addition we aim to provide predictions on feasibility of such joint measurements in the near future.}
   { We study which GW observables can be combined with measurements from gamma--ray polarimetry to improve the discriminating power regarding GRB emission models. We then provide forecasts for the joint detection capabilities of current and future GW detectors and polarimeters.}
   {Our results show that by adding GW data to polarimetry, a single precise joint detection would allow to rule out the majority of emission models. We show that in the coming years joint detections between GW and gamma--ray polarimeters might already be possible. Although these would allow to constrain part of the model space, the probability of highly constraining joint detections will remain small in the near future. However, the scientific merit held by even a single such measurement makes it important to pursue such an endeavour. Furthermore, we show that using the next generation of GW detectors, such as the Einstein Telescope, joint detections for which GW data can better complement the polarization data become possible.}
   {}


\maketitle

\section{Introduction}

Gamma--Ray Bursts (GRBs) are the brightest electromagnetic events in the Universe since the Big Bang. Discovered first in 1967, they remain one of the most researched topics in astrophysics. These phenomena consist of a gamma--ray component lasting from seconds to minutes, referred to as the prompt emission, followed by a longer lasting afterglow observable over a wide range of the electromagnetic spectrum. Since their discovery much has been learned about GRBs. Firstly, through measuring the location of these transients in the sky, they were found to have an extra--galactic nature \citep{Meegan:1992xg, Costa:1997obd, vanParadijs:1997wr, 1997Natur.387..878M}. In addition, measurements of the light curves (the intensity of the gamma--ray emission over time) allowed to divide GRBs into 2 classes \citep{Kouveliotou:1993yx}: long and short GRBs. Long GRBs, classified as those with a duration exceeding 2 seconds, have been associated with the death of massive stars \citep{Galama,Patat}, thanks to joint observations with X--ray, optical and radio waves \citep{Cano:2016ccp}. The prompt radiation from the death of the massive star is theorized be be emitted from two jets powered by a central engine. Interactions of the jet with the interstellar medium subsequently result in the afterglow. Short GRBs, which emit their radiation in a similar way, were long theorized to be the result of the merger of two compact objects such as binary neutron stars \citep{Eichler:1989ve}. Further evidence of this was found in 2017, through the joint detection of GRB 170817A with the Gravitational Wave event GW170817. Despite the wealth of knowledge gained on GRBs over the decades, much remains still unknown about these extreme phenomena. Some of the most important open questions concern the jet structure, the importance of magnetic fields, as well as the physical mechanism responsible for the gamma--ray emission. Some of these questions can be resolved through GW observations (see for example \cite{LIGOScientific:2011cgo} for an overview), while others can be answered through polarization measurements of the prompt emission \citep{Gill:2021jev}. 

The value of GW astronomy to GRB physics became clear with the observation of GW170817. This GW was significantly detected by both LIGO sites (with \ac{SNR} of \VAR{gw170817.SNR.LIGO_H|round(1)} and
\VAR{gw170817.SNR.LIGO_L|round(1)} respectively). In addition, Virgo recorded a marginal signal with a SNR of \VAR{gw170817.SNR.Virgo|round(1)} and did significantly contribute to the localization of the source \citep{LIGOScientific:2017vwq}.  The measurements do allow to constrain the inclination angle of the binary with respect to the line of sight,  $\iota=147\substack{+23 \\ -27}~\si{\degfull}$. The second angle defining the orbital plane, $\psi$ (see \autoref{sec:gw_parameter} for a detailed definition), which is of particular interest for the work presented in this paper, could however not be constrained. This is due to a combination of the low sensitivity and the orientation of the detectors. 

Apart from a measurement in GW, the event was observed by the \textit{Fermi}-Gamma Ray Burst Monitor (GBM) as well as the INTEGRAL SPI-ACS gamma--ray detectors \citep{LIGOScientific:2017ync}. This confirmed the connection between a binary neutron star merger event and the GRBs. The prompt gamma--ray emission was followed by detailed observations of the GRB afterglow in X--ray, optical and radio waves \citep{LIGOScientific:2017ync,Coulter:2017wya, Smartt:2017fuw, Pian:2017gtc}. Amongst the many novel measurements made on the system, these measurements allowed to locate the host galaxy and thus provide the sky location and redshift of the event. This provided additional constrains on $\iota$, resulting in  $\iota=147\substack{+13 \\ -10}~\si{\degfull}$ (or, using the convention from \cite{Finstad:2018wid}, $\Theta=32\substack{+10 \\ -13}~\si{\degfull}$, with $\Theta=\min\{\iota, \SI{180}{\degree} -\iota\}$). Unfortunately the knowledge of the host galaxy, while improving the measurement of $\iota$ because of its degeneracy with the luminosity distance, does not result in any significant improvement in the determination of $\psi$, which is degenerate with other parameters, as can be seen from the posterior samples published together with \cite{Finstad:2018wid}.  In addition to the detection of the host galaxy, radio follow--up measurements resulted in the detection of the super--luminal movement of the shock induced by the jet interaction with the interstellar medium \citep{Mooley:2018qfh, Ghirlanda:2018uyx}. Such radio measurements will allow to contribute further to the constraint on the prompt emission models, as will be discussed later in this work. 

The detection of GW170817 was followed during the first part of the third observing run by GW190425, an additional binary neutron star (BNS) merger \citep{LIGOScientific:2020aai}, and during the second half of O3 by the two neutron star--black hole (NSBH) events GW200105 and GW200115 \citep{LIGOScientific:2021qlt, LIGOScientific:2021psn}. None of these detections were, however, accompanied by GRB observations, something which can - at least in part - be attributed to the narrow opening angle of the GRB emission. For short GRBs where the opening angle was measured using the jet break time, opening angles of $6.1\substack{+9.3 \\ -3.2}$ degrees are found \citep{Alicia}. When including lower limits on the opening angle, calculated using the last observation time of the afterglow, opening angles are constrained to approximately $\SI{16}{\degree}$ for short GRBs \citep{Fong:2015oha,Salafia:2022dkz}, while studies in \cite{Alicia} also show evidence for wide opening angles, up to $34^\circ$ in one case. The coming decade will allow for an increasing probability for joint GW--EM detections, thanks to the increasing sensitivity of GW detectors and the construction of new facilities. In chronological order, the first GW network configuration we will consider is the fifth observing run (O5) of second generation detectors, i.e. LIGO, Virgo and KAGRA, which will potentially be joined by LIGO India, currently under construction. O5 is expected to begin in 2026, and last for about 2 years. After this, the LIGO detectors are expected to undergo major upgrades to reach the so--called Voyager stage, with a sensitivity about two times better than the Advanced+ design \citep{LIGO:2020xsf}. The next step are then third--generation detectors, namely Einstein Telescope (ET) in Europe \citep{Punturo:2010zz, Hild:2010id} and Cosmic Explorer (CE) in the U.S. \citep{Reitze:2019iox, Evans:2021gyd}, which are expected to start their operations in the middle of the next decade. 

In parallel to the increasing sensitivity of the GW detectors several gamma--ray detector constellations are planned to be launched in the coming years. Such constellations are specifically aimed at providing detailed GRB spectral and location measurements. Some examples of these are CAMELOT \citep{CAMELOT}, HERMES \citep{HERMES-SP:2021hvq} and GRID \citep{Wen:2019rxw}. In addition, two large scale gamma--ray polarimeters are planned. Polarimetry of the prompt emission has long been theorized to be a unique probe of GRB physics allowing to answer a range of open questions which appear to be impossible to answer using spectral measurements alone \citep{Gill:2021jev}. Polarization measurements on GRBs allow, for example, to constrain the emission mechanisms at play, and provide information on the emission region, such as the presence of magnetic fields and their orientation \citep{toma2009}. These measurements provide both the Polarization Degree (PD), defined as the percentage of the gamma--rays polarized in a non--random direction, and the Polarization Angle (PA), the preferred angle of the polarization vectors of the gamma--rays as measured against the north celestial pole (galactic north). Due to the wealth of information which can be probed using polarization measurements, attempts have been made to measure it over the last 25 years using a range of missions \citep{Gill:2021jev}. While initial attempts were plagued by systematic issues and limited statistics, thereby not allowing to constrain the polarization parameters, the first generation of dedicated polarimeters launched in the last decade has seen more success. While high levels of polarization have been reported, see e.g. \cite{GAP4} which reports PDs for 2 GRBs exceeding $70\%$ and \cite{2013MNRAS.431.3550G} which reports a PD exceeding $60\%$, GRB polarization catalogs by POLAR \citep{Kole:2020eef}, which presents measurements for 14 GRBs, and AstroSAT-CZTI \citep{Chattopadhyay:2022yzx}, which presents measurements for 20 GRBs, find low levels of polarization (see \cite{Gill:2021jev} for a detailed review). The low levels of polarization  further push the need for dedicated large scale instruments with sufficient sensitivity to properly probe the polarization parameters. The second generation of dedicated GRB polarimeters consists of LEAP, POLAR-2 and potentially the Dhaksa mission.\footnote{\url{https://www.star-iitb.in/research/daksha}.} LEAP \citep{10.1117/12.2594737} is in the proposal phase, currently in a phase A study, for a NASA  launch in 2027. POLAR-2 has been accepted for launch in 2025 towards the China Space Station (CSS) \citep{POLAR-2:2021uea}. Both instruments have a large effective area which is required to achieve a large enough signal to perform significant polarization measurements. Thanks to these effective areas, which are foreseen to be exceed $1000\,\mathrm{cm^2}$, and a field of view (FoV) of approximately half the sky, these instruments will prove to be extremely valuable in the search for GW counterparts. 

Apart from increasing the probability for joint detections with mergers of compact binaries, these polarimeters will allow to answer questions regarding the origin of the gamma--ray emission by measuring the polarization of GRBs, as outlined for example in \cite{Gill:2021jev}. Individual measurements of the PD and PA of the gamma--ray prompt emission will not suffice to answer such questions as the exact PD predicted by emission models depends on several unknown parameters which vary from GRB to GRB. This results in a relatively large and overlapping range of the PD values predicted by the various models. In order for polarimeters to answer the questions regarding the emission mechanisms and environments they therefore need to measure the properties for a large set of GRBs, thereby producing a distribution of PD values which can be compared to theoretical models. As we will demonstrate in this paper, this situation changes when such polarization measurements are combined with information from GW measurements. GW measurements will allow to constrain the orientation of the progenitor for short GRBs. This orientation is correlated to the polarization predicted by various models. Therefore, measuring the orientation will allow to constrain the theoretical predictions on PD and PA, thereby giving a single joint detection the potential to discriminate between emission models. 

In order to explore the potential of joint GW and gamma--ray polarization measurements, we will first present the geometry of the binary system, followed by a description on how the various angles defining the system can be measured using GW detectors in \autoref{sec:gw_parameter}. In \autoref{sec:GRB_pol} we present qualitative predictions on the PD and PA values by various models and describe the correlation between them and the two binary orientation angles $\iota$ and $\psi$. Finally, in \autoref{sec:future_prospects} we will present predictions for the joint observation probability in the coming decade and finish with a conclusion in \autoref{sec:conlcusions}.

\section{Parameterising and inferring the geometry of a binary system from GW emission}\label{sec:gw_parameter}
In the most general case of quasi--circular orbits, the response of a GW detector to the emission of a coalescing binary system can be parameterised in terms of 17 parameters (for a pedagogical introduction see e.g. \cite{Maggiore:2007ulw}), $\vb*{\theta} = \{m_1, m_2, d_L, \chi_{1,x}, \chi_{2,x}, \chi_{1,y}, \chi_{2,y}, \chi_{1,z}, \chi_{2,z}, \Lambda_1, \Lambda_2, \theta, \phi, t_c, \Phi_c, \iota, \psi, \}$ where $m_1$ and $m_2$ denote the detector--frame masses of the two objects, $d_L$ the luminosity distance to the source, $\chi_{i,c}$ the dimensionless spin of the object $i=\{1,2\}$ along the axis $c = \{x,y,z\}$, $\Lambda_i$ the dimensionless tidal deformability of the object $i$ (vanishing in the case of a BH), $\theta$ and $\phi$ are the sky position coordinates, $t_c$ the time of coalescence, and $\Phi_c$ the phase at coalescence. The remaining parameters are the ones characterising the binary geometry we are interested in: the inclination angle of the binary with respect to the line of sight, $\iota$, and the polarization angle, $\psi$. 
The explicit expression of the Fourier--domain signal as a function of these parameters is given by
\begin{equation}\label{hTime}
\tilde{h}(f;\vb*{\theta},\vb*{\lambda}) = \tilde{h}_{+}(f;\vb*{\theta})F_{+}(f; \vb*{\theta}, \vb*{\lambda}) + \tilde{h}_{\times}(f;\vb*{\theta}) F_{\times}(f; \vb*{\theta}, \vb*{\lambda})\, ,
\end{equation}
where $\vb*{\lambda}$ are the parameters characterising the detector (i.e. its latitude, longitude, orientation and shape), $\tilde{h}_{+,\times}$ are the two polarizations of the GW signal and $(F_+, F_{\times})$ the so--called ‘‘antenna pattern functions'', needed to project the signal onto the detector arms.

\begin{figure}
  \centering
  \includegraphics[width=0.5\linewidth]{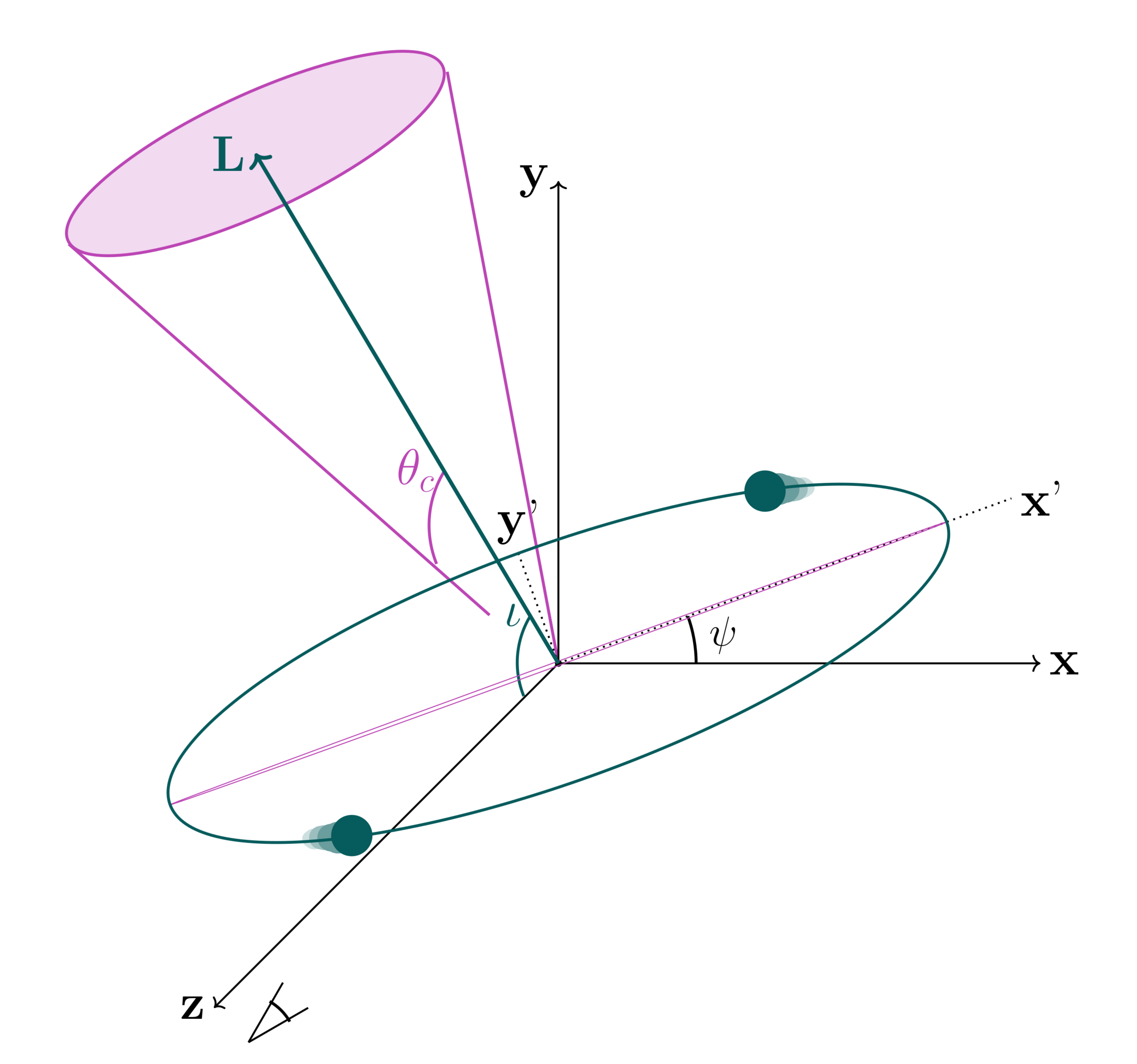}
  \caption{Schematic representation of the binary system merger. A coordinate system (x,y,z) is shown, where the line of sight coincides with the z--axis. The primed coordinate system is achieved through a rotation by angle $\psi$ in the x--y plane and with an angle $\iota$ around the x--axis. Through this the jet with angular momentum vector $\vec{L}$, is aligned with the the z' axis. The jet plane, from which the $\gamma$--rays are emitted, is indicated in purple while the half jet opening angle $\theta_c$ is shown as well. In the current figure the $\gamma$--ray emission would not be observed, as the jet plane points away from the observer.}
  \label{fig:gw_angles}

\end{figure}

It is interesting to briefly discuss the physical meaning of $\iota$ and $\psi$, which can be understood from \autoref{fig:gw_angles}. When performing the computation of the GW signal emitted by a coalescing binary, one has to choose the coordinates in the frame of the source (the radiation frame), and then relate them to the observer frame. The standard practice is to choose one of the axes to coincide with the binary orbital angular momentum, whose inclination with respect to the direction of observation is the angle $\iota$.\footnote{Notice that, in principle, one would have to consider the contribution of unaligned spin components, which cause orbital precession (in this case, the inclination is defined with respect to the the total angular momentum rather than the orbital angular momentum, and the variable commonly denoted as $\theta_{JN}$, which equals $\iota$ in the non--precessing case). While their inclusion could in principle help in disentangling the two polarizations, and thus in measuring the inclination angle and $\psi$, given the small observed adimensional spins of neutron stars in binaries, of $\order{0.05}$ in magnitude, this effect is found to be negligible.} The remaining two axes define the polarization tensors of the wave, and will in general be rotated with respect to the corresponding axes in the observer frame by an angle $\psi$ (see e.g. Sec. 1 and 7 of \cite{Maggiore:2007ulw} and \cite{Isi:2022mbx}), which is consequently called \emph{polarization angle} (not to be confused with the polarization angle of the gamma--rays which we refer to as PA).

Accounting for the effect of Earth motion, which is relevant for a 3G detector, where a signal can stay in band up to $\order{1~{\rm day}}$, the two polarizations of the Fourier--domain signal for the dominant quadrupole mode\footnote{While the contribution of higher--order modes can help in disentangling the two polarizations of the wave, its impact for BNS signals is negligible, given the small masses of the objects and the narrow range for the mass ratio. In the following we will thus consider the fundamental mode only, adopting for the analysis the \texttt{IMRPhenomD\_NRTidalv2} waveform model \citep{Khan:2015jqa, PhysRevD.100.044003}, which includes the contribution of tidal effects.} depend on $\iota$ and $\psi$ as 
\begin{subequations}\label{eq:hphc_expr}
    \begin{align}
        &\tilde{h}_{+} (f;\vb*{\theta}, \vb*{\lambda}) = \left[a(f;\vb*{\theta}, \vb*{\lambda}) \cos2\psi + b(f;\vb*{\theta}, \vb*{\lambda})\sin2\psi\right] \, A^{(F)}(f; \vb*{\theta}) \dfrac{1+\cos^2\iota}{2}\, e^{i\Phi(f;\vb*{\theta}, \vb*{\lambda})}\,,\\
        &\tilde{h}_{\times} (f;\vb*{\theta}, \vb*{\lambda}) = \left[b(f;\vb*{\theta}, \vb*{\lambda}) \cos2\psi - a(f;\vb*{\theta}, \vb*{\lambda})\sin2\psi\right]\, A^{(F)}(f; \vb*{\theta})\, (\cos\iota)\, e^{i [\Phi(f;\vb*{\theta}, \vb*{\lambda}) + \nicefrac{\pi}{2}]}\,,
    \end{align}
\end{subequations}
where the GW phase is given by
\begin{equation}
    \Phi(f;\vb*{\theta}, \vb*{\lambda}) = 2\pi f t_c +  \phi_{L}(f; \vb*{\theta}, \vb*{\lambda}) - \Phi_c + \Phi^{(F)}(f; \vb*{\theta}) -\frac{\pi}{4}\,,
\end{equation}
with $\phi_L$ being the location phase (due to the time difference between the geocentric and the detector reference frames), $\Phi^{(F)}$ and  $A^{(F)}$ the output phase and amplitude of the waveform model, respectively, and the projection functions $a(f; \vb*{\theta}, \vb*{\lambda})$ and $b(f; \vb*{\theta}, \vb*{\lambda})$ encode the effect of the rotation of the Earth and are given in Eq. 17 of \cite{Iacovelli:2022bbs}. Notice that, as it is apparent from \autoref{eq:hphc_expr}, the GW signal actually depends on $2\psi$, thus it is not possible to distinguish between two system that differ by a \SI{180}{\degree} rotation on the orbital plane through the emitted gravitational radiation. As the polarization of the gamma--rays exhibits the same rotational symmetry, this will not affect the discussion later on in this work. 

To observe the EM counterpart of a system, being the emission beamed, the inclination angle should be smaller or close to $\theta_c$, the half jet opening angle of the GRB. Although emission is still expected for $\iota > \theta_c$, as was likely the case for GRB 170817A, the fluence of the EM part will rapidly decrease with increasing $\iota$. As is apparent from the above equations, for optimally oriented systems (i.e. with $\cos\iota\sim1$) the two GW polarizations cannot be properly disentangled, leading to a degeneracy between $\psi$ and other parameters ($\Phi_c$ in particular), which degrades the measurement of this angle. 

\section{GRB Prompt Polarization}\label{sec:GRB_pol}

\subsection{\texorpdfstring{$\gamma$}{gamma}--ray Polarimetry}

Despite thousands of GRB detections, many open questions remain regarding the origin of the high energy prompt emission. For example, no consensus exists regarding the composition, structure and emission mechanisms at play within the relativistic jets thought to be responsible for the $\gamma$-ray emission \citep{Gill:2021jev}. The answer to many of these open questions are theorized to be embedded in the polarization of the prompt emission, as different emission models predict different PDs \citep{toma2009}. The gamma--rays emitted from the jet, can be polarized when emitted e.g. through synchrotron emission which would indicate the presence of magnetic fields. When the dominant emission mechanism is inverse--Compton scattering, indicating that magnetic fields are not playing an important role, the emission is generally unpolarized \citep{Lundman:2013qba}. It should be noted that in this work we will present only a qualitative description of the polarization properties in order to keep this work relatively concise and accessible to the wider multi--messenger community. Although various emission models will be discussed here, within them a number of free parameters exist which affect the precise value of PD, and in some cases PA. Providing detailed quantitative predictions for each model would require assumption on the various unknown model details, such as the still unknown jet structure. While the importance of such issues will be discussed, the reader is referred to overview papers such as \cite{Gill:2021jev} and \cite{toma2009} for details.

While the first polarimetry measurement attempts date from the start of this century (see \cite{Gill:2021jev} or \cite{McConnell:2016lwd} for a detailed overview), such measurements were often plagued by systematic errors as well as by limited statistical significance. Such issues motivated the development of instrumentation fully dedicated to GRB polarization measurements in the form of GAP \citep{Yonetoku:2011zz} and POLAR \citep{Kole:2022gxs}. In addition, measurements by non--dedicated instruments whose polarization measurement capabilities were shown prior to launch (such as the CZTI on AstroSAT \citep{Chattopadhyay:2022yzx}) were performed in the last decade. The results produced by such instruments have shown an increase in precision, albeit not yet at a level allowing for discrimination between the various existing emission models. This is in part a result of the difficulty in these measurements, while the generally low levels of polarization observed by these instruments increase the difficulty in the measurements as well \citep{Gill:2021jev}. 

Despite the lack of measurements capable of answering some of the remaining open questions regarding the nature of GRBs, the significant progress made in recent years in $\gamma$-ray polarimetry has inspired the development of the next generation of polarimeters in the form of POLAR-2 and LEAP. As will be shown in \autoref{sec:future_prospects}, their sensitivity will allow these instruments to measure the polarization of a large number of GRBs, including those only marginally brighter than GRB 170817A. The high precision in PD measurements for bright GRBs, and the generally large number of GRBs for which constraining PD measurements are possible, will allow these polarimeters to start probing the various emission models. Probing them will, however, still require a large set of measurements as the PD and PA predicted by the various existing models span a large range. This is a result of the dependence of the PD and PA on a significant number of variables that are either unknown or currently impossible to measure. For example, the PD will depend on the jet profile of the Lorentz factor inside of the jets, something which remains poorly understood. In addition, the PD and PA, as we will show here, will be highly dependent on the orientation of the progenitor system with respect to the observer. As a result, the so--called photospheric emission models, which will be discussed in \autoref{sec:GRB_pol}, typically predict unpolarized emission; however, PDs of up to $40\%$ can still be achieved for specific viewing angles \citep{Lundman:2013qba}. As a PD of $40\%$ is also possible to produce with competing models, the measurement of a single GRB with a PD of $40\%$ does not suffice to discriminate between models. Rather the measurement of a distribution of a large number of GRBs is needed. As GW measurements are able to constrain several parameters unavailable using traditional astronomy, it is of interest to investigate how such parameters can be combined with polarimetry measurements to increase their individual intrinsic scientific potential. For this purpose, we will first provide a qualitative description of the way in which GRB polarimetry can distinguish between various $\gamma$-ray emission models and then discuss the complementary nature of GW measurements. Despite not providing quantitative predictions for all the possible different parameter spaces within the presented models, the general correlation between the GW parameters and the PD and PA are largely independent on these model details. Therefore, a qualitative description is sufficient here, while detailed predictions will be the subject of future works.

\subsection{Synchrotron vs. Photospheric Emission}

One of the most debated questions regarding GRB physics is the mechanism responsible for the $\gamma$-ray emission. Although a large variety of models persists to this day, all of these either attribute the emission to (inverse--)Compton scattering or synchrotron emission. The non--thermal spectrum of the prompt emission, which has been observed in 1000's of GRBs to date, is typically fit using the empirical Band--function \citep{Band+93} which consists of a smoothly broken power--law. Despite  several observed discrepancies among the measured spectral shape and the one predicted by the two classes of models, in general both of them can be used to predict the emitted spectrum with sufficient accuracy. The models invoking synchrotron emission aim to explain the observed spectra through optically--thin synchrotron emission produced by relativistic electrons. These electrons are accelerated at internal shocks resulting from the collision of baryonic shells in the emission jets of the GRB. Over the years one issue which was found with such models is the observed spectral slope below the break in the spectrum. The steepness of this slope, observed in several GRBs, according to some of the first synchrotron models could not exceed $-1.5$ \citep{Preece1998}. GRBs for which fits with the Band function yielded values exceeding this limit, often referred to as the synchrotron line of death (see for example \cite{Preece1998}), prompted alternative models to be considered. The most prominent alternative models use a process with multiple inverse--Compton scatterings on sub--relativistic electrons to explain the observed emission spectrum (see for example \cite{2000ApJ...530..292M,2011ApJ...737...68B}). These models are capable of explaining the Band--like spectrum and are referred to as dissipative photospheric models. 

It should be noted that, in recent years, thanks to advances in the analysis methods, in which fitting with the empirical Band function was replaced by fitting physical models to the data, the initial concerns regarding synchrotron emission have been removed (see e.g. \cite{Ravasio:2017vak, Burgess:2018dhc, Ravasio:2019kiw, Oganesyan:2019fpa, Zhang:2020ccp}). As a result, both synchrotron emission and  photospheric models are currently capable of explaining the observed spectral features. Additionally, some alternatives have been proposed, such as the Compton drag model \citep{Dar:2003vf} that was used to explain the high levels of polarization measured in GRBs, which were later found to be incorrect \citep{Lazzati:2003af}. Although in this work we will provide a qualitative description of the polarization predictions of photospheric and synchrotron emission only, the Compton drag model, which also relies on inverse Compton scattering, is foreseen to show similar behaviour to the photospheric models albeit with typically higher levels of polarization \citep{Gill:2021jev}. 

\subsection{Polarization Degree Predictions}\label{sec:intro_PDP}

As the spectral and timing features of GRBs generally match the predictions from both photospheric and synchrotron models, another observable of the prompt emission is required to distinguish between them. For this purpose, good candidates have been theorized to be the polarization parameters, primarily the PD. As we will show below, the PD predicted by photospheric models is typically low, whereas synchrotron emission can produce higher levels of polarization depending on the magnetic field configuration. Typically, however, the PD predicted by most models exhibits a low level of polarization for on--axis observations. To allow for discrimination between such models a large sample of observed GRBs are needed in order to study the distribution of the PD and, for example, its dependence on the energy of the spectral break. Regarding the Polarization Angle (PA), the predictions by synchrotron and photospheric emission differ. However, as the GRB system cannot be spatially resolved using gamma--rays, its geometry is unknown, thereby not allowing to relate the PA to any axis in the system. As a result, the absolute PA of the emission, although always reported, is not considered as a parameter with discriminative power Note, however, that a time evolution of the PA during the emission has been reported by several groups, see for example \cite{Gotz:2009ak,GAP2,Burgess:2019tjv} and \cite{Sharma:2019nfc}\footnote{The significance reported in this work was found to be overestimated in \citep{Chattopadhyay:2022yzx}}. The exact nature of the evolution, whether it is for example a $90^\circ$ flip or a smooth evolution, cannot be determined with the existing data, making it currently difficult to interpret. Although the time evolution of the PA could potentially be used to discriminate between models, the time averaged PA cannot. As we will show in the following, by combining GRB and GW measurements, the PA can be related to $\psi$, thus potentially becoming a powerful probe in GRB physics.

\subsubsection{Photospheric Emission}

The physical mechanism through which the high energy emission in a GRB jet is produced in photospheric emission models is (inverse--)Compton scattering. The cross section for a Compton scattering photon is defined by the Klein--Nishina equation

\begin{equation}\label{eq:KN}
    \frac{d\sigma}{d\Omega} = \frac{r_o^2}{2}\frac{E'^2}{E^2}\left(\frac{E'}{E}+\frac{E}{E'}-2\sin^2\theta_\gamma \cos^2\phi_\gamma\right).
\end{equation}
Here $r_0 = e^2/m_ec^2$ is the classical electron radius, with $e$ being the elementary charge, $E$ is the initial photon energy, $E'$ the final photon energy, $\theta_\gamma$ the polar scattering angle, and $\phi_\gamma$ the azimuthal scattering angle as measured with respect to the polarization vector of the incoming photon. In the case of Thomson scattering, where $E'=E$, the cross section has a maximum dependence on $\phi_\gamma$ when $\theta_\gamma=\SI{90}{\degree}$ while being independent on $\phi_\gamma$ when $\theta_\gamma=\SI{0}{\degree}$, as can be easily deduced from \eqref{eq:KN}. As a result, the polarization of an initially unpolarized beam can increase up to $100\%$ when selecting only the emission which underwent scattering at $\theta_\gamma=\pi/2$. For forward scattering photons, however, the flux would remain unpolarized. In the Compton scattering regime, the PD for $\theta_\gamma\sim\SI{90}{\degree}$ decreases but remains significant up to $\gamma$--ray energies. Assuming an initially unpolarized flux, coming for example from thermal emission produced in an optically thick outflow in the GRB jet, the observed photons scatter while traveling. Assuming a spherical expansion, the polarization angle is randomized at each scattering interaction, until the photon escapes the medium. This happens around the photosphere, where the medium becomes optically thin. %

\begin{figure}
  \centering
  \includegraphics[width=0.5\linewidth]{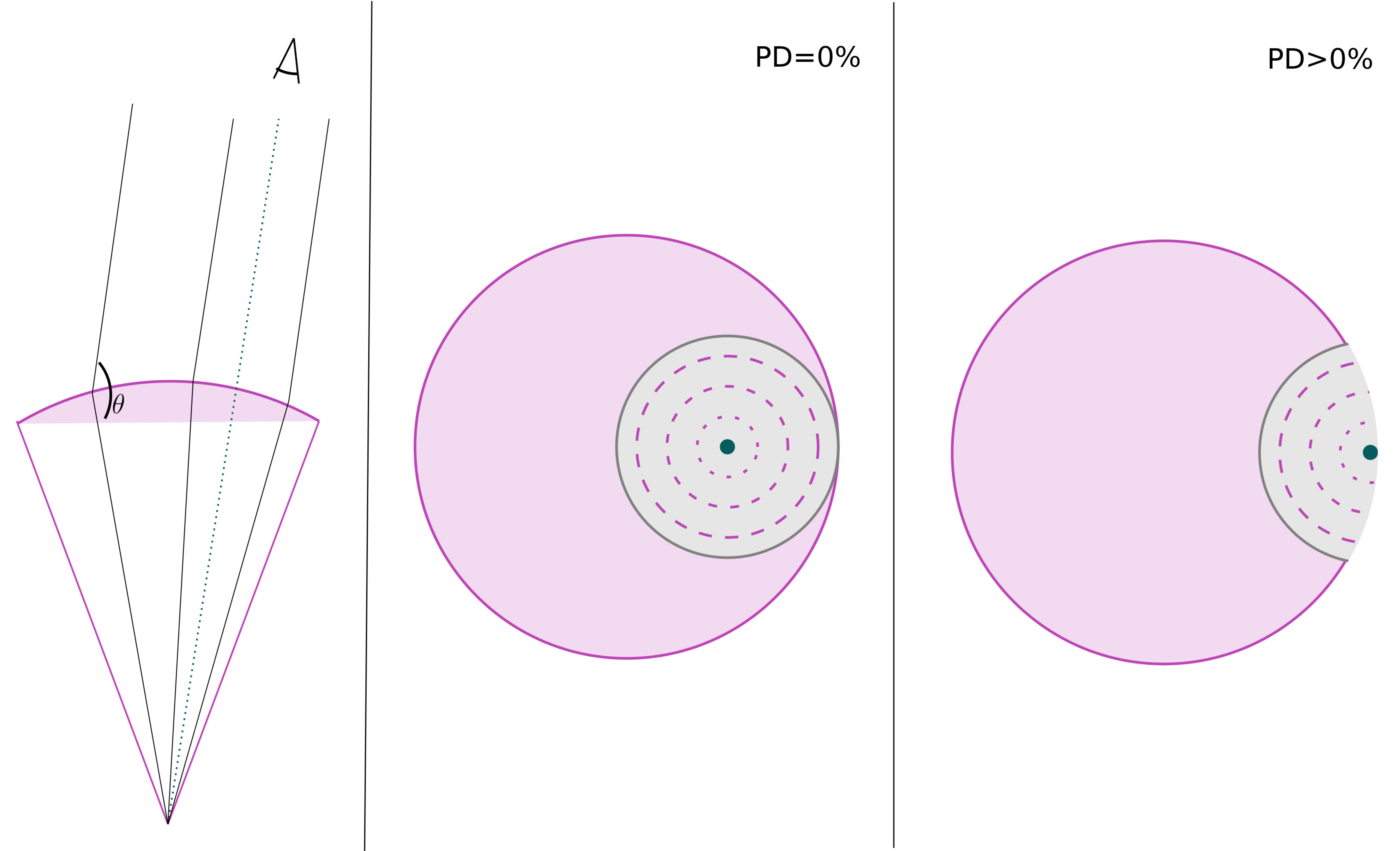}
  \caption{Schematic illustration of the polarization from photospheric emission. \textit{Left}: Illustration of the photons (black lines) scattering from the jet (purple) towards the observer. The green dotted line represents the observers viewing axis. The photons scattering towards the observer in the photosphere (top layer of the jet), close to the line of sight will have a scattering angle $\theta\approx \SI{180}{\degree}$, whereas $\theta$ goes closer to $\SI{90}{\degree}$ when moving away from the viewing axis. As a result, the polarization increases with the polarization angle into the drawn plane. \textit{Middle}: The jet (purple) seen from the top along with the observed area (gray, with size $1/\Gamma$) centred around the viewing axis (green). With the full observed area contained in the jet, the various polarized areas will cancel out one another, resulting in no net polarization. \textit{Right}: Same as the middle panel in a situation in which the observer area not fully contained in the jet, allowing for a net observed polarization, with a polarization angle perpendicular to the off--axis viewing angle ( $\theta_v=\iota$).}
  \label{fig:photo_jet}
\end{figure}

In order to achieve polarized emission, an anisotropy in the scattering direction of the photons in the medium is required. Such an anisotropy can be achieved by beaming of the relativistic outflow, such as in the case of a jet. The polarization of the photons emitted through inverse--Compton scattering in the photospheric models is related only to their last scattering interaction. The PA of the photon will be perpendicular to the plane defined by its incoming and outgoing direction. For photons emitted along the line of sight towards the observer $\theta_\gamma\simeq\SI{0}{\degree}$, resulting in a negligible PD. As illustrated in \autoref{fig:photo_jet}, the polarization increases for photons emitted further away from the line of sight since these will have undergone on average a larger polar scattering angle. As indicated in \autoref{fig:photo_jet}, this results in a polarization angle perpendicular to the radial axis along the line of sight (in the left panel the angle is perpendicular to the plane of the image). Despite the significant polarization in parts of the jet, an observation of the full jet will still lead to a net polarization of $0\%$, as the distribution of the polarization vectors is symmetric, resulting in a full cancellation. However, due to the relativistic beaming, the observer will only see an area with size $\Gamma^{-1}$, where $\Gamma$ is the Lorentz factor of the emission medium. This effect, in combination with larger viewing angles, allows to break the symmetry of the observation, thereby allowing to observe a net polarization. 

A significant polarization can be observed if the viewing angle with respect to the jet center $\theta_v$ (where $\theta_v\sim\iota$) exceeds $\theta_c$. From this somewhat simplified qualitative description, it is clear that polarization will only be observed for off--axis GRBs. In addition, as the preferred scattering angle required for photons to reach the observer is larger for areas further from the viewing axis, the net PD is larger as $\theta_v/\theta_c$ increases. As a result, the PD increases with off--axis viewing angle, and therefore with $\iota$.

It should be noted that, although providing the correct correlation between $\iota$ and the PD, the description of the emission provided here lacks some of the complex details, which can be found e.g. in \cite{Lundman2014, Lundman2018, Gill2018}. Such details allow us to calculate the exact PD, as well as its exact dependency on $\iota$. As such quantitative results strongly depend on a significant number of unknown parameters, related to the jet structure, it is not possible to provide a direct quantitative prediction without making a number of assumptions. Details of these unknown parameters are discussed in depth in for example \cite{Gill2018}, which shows that the levels of observed PD can be as high as $15\%$ for large off--axis observations. This value, however, depends on both the size of the jet, the Lorentz factor and its angular structure. In the simplest model, the jet consists of an outflow with a uniform Lorentz factor, which drops off sharply beyond $\theta_{c}$. This is called a top--hat jet. In a structured jet, a layer of sheared material is added beyond $\theta_{c}$, whose Lorentz factor drops off as $\theta$ (the angle with respect to the jet symmetry axis) goes beyond $\theta_{c}$. In \cite{Lundman2014, Ito2013} Monte Carlo simulations of a structured jet with a sharp drop off in Lorentz factor in the shear layer are shown to give levels of polarization exceeding $20\%$. Apart from the structure of the jet, the radial distance between the last interactions of the photons also plays an important role, as it affects the mixing of photons emitted from various locations that reach the observer. The observed PD therefore also depends on the density profile around the photosphere. Generally, however, all the models produce the same qualitative prediction of an increasing PD with increasing $\iota$.

\subsubsection{Synchrotron Emission}

\begin{figure}
  \centering
  \includegraphics[width=0.5\linewidth]{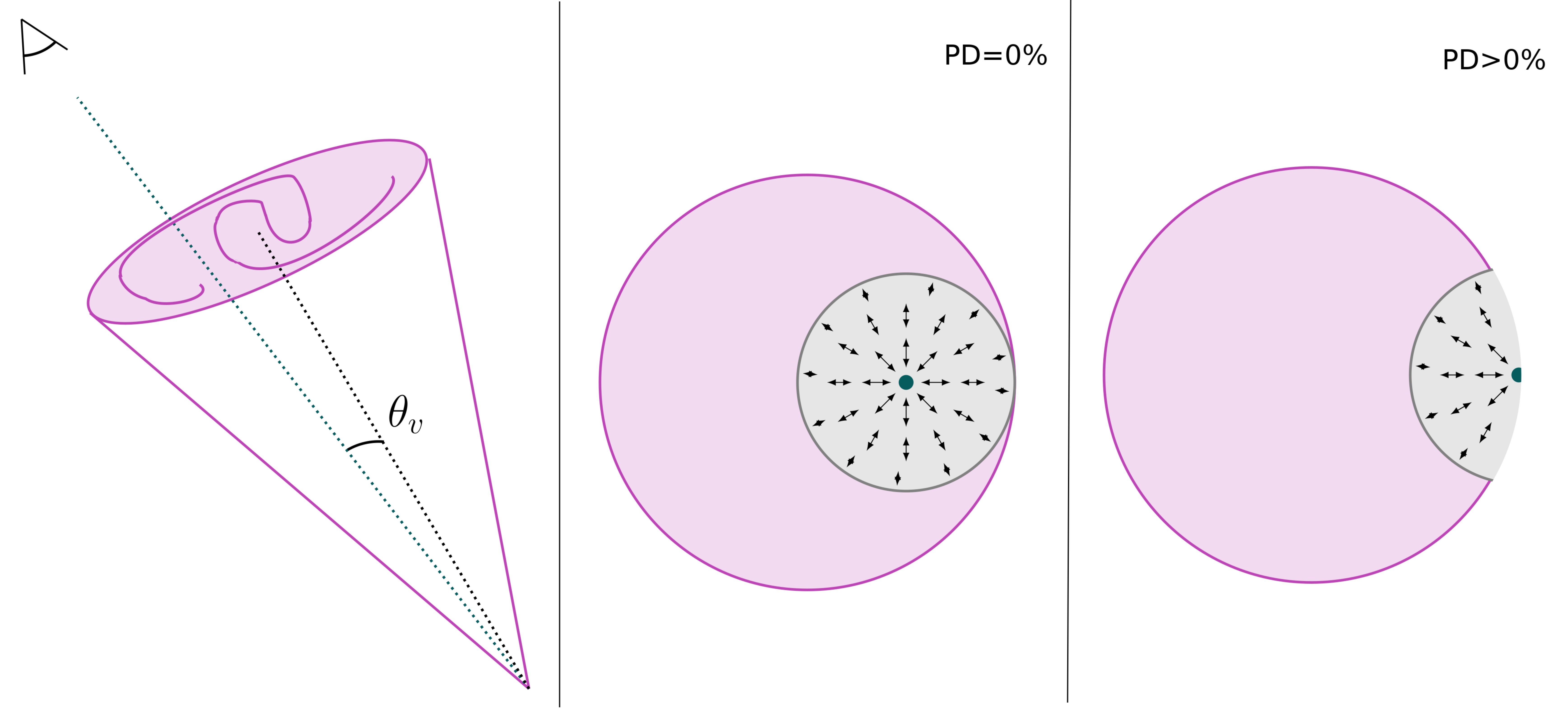}
  \caption{Schematic illustration of the polarization from synchrotron emission from a random magnetic field. \textit{Left}: Illustration of the emission of the photons towards an observer with a line of sight forming a viewing angle $\theta_v$ with the jet. The emission stems from magnetic field lines with a randomized direction within the ejection plane, that are indicated with the purple line. \textit{Middle}: The observer sees the emission area with a size of $1/\Gamma$ (shown in gray) within the jet (shown in purple). The polarization vectors of the photons are indicated as black arrows with the direction indicating the polarization angle and the size representing the PD. For an observation area contained within the jet, the PD is $0\%$ due to cancelling out of the polarization vectors within this area. \textit{Right}: As in the middle panel, in a situation in which the observed area is not fully contained in the jet, allowing for a net observed polarization, with a polarization angle aligned to the off--axis viewing angle ( $\theta_v=\iota$).}
  \label{fig:synch_random_jet}
\end{figure}

The PD of emission from synchrotron models can be as high as $70\%$ \citep{Covino2016}. The collisionless shocks formed in the jet may produce sizable magnetic fields with random directions. In case there is some level of alignment in the magnetic field directions within the jet plane, e.g. as it is often assumed in the propagation direction of the jet, significant levels of polarization can be achieved. In such synchrotron models, often referred to as synchrotron random--field models (SR), when assuming a magnetic field along the jet direction, the polarization angle of the flux will be perpendicular to the radial vector of the jet, resulting in a rotational symmetry \citep{toma2009,Gill2018}. Although the PA differs from that of photospheric emission, here again the observed PD will be 0\% when the full emitting area is observed due to the symmetry in the PA. A significant PD can be observed when $\theta_v>\theta_c$. The observed area, with a size of $\Gamma^{-1}$ within the overall jet is illustrated in grey in \autoref{fig:synch_random_jet}. It can be observed that the PD will show a similar behaviour to the one described for photospheric emission. Only when the symmetry is broken, which means $\theta_v > \theta_c$, the observer will measure a non--zero PD.

\begin{figure}
  \centering
  \includegraphics[width=0.5\linewidth]{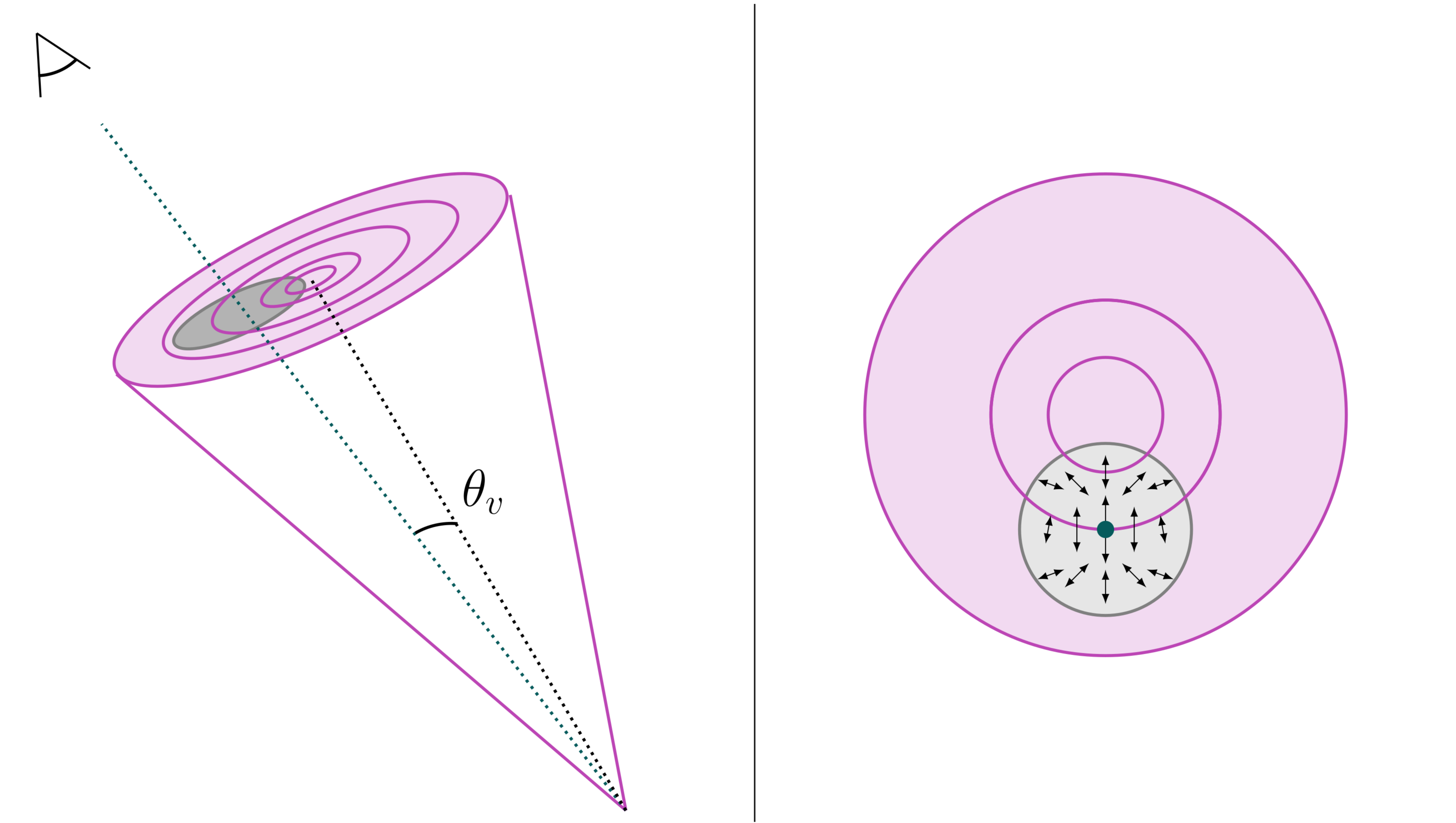}
  \caption{Schematic illustration of the polarization from synchrotron emission from a toroidal magnetic field. \textit{Left}: Illustration of the emission of the photons towards an observer  with a line of sight forming a viewing angle $\theta_v$ with the jet. The emission stems from a magnetic field whose lines are indicated in purple. The gray area corresponds to the area (with a size of $1/\Gamma$) which is seen by the observer. \textit{Right}: The observer sees the emission area with a size of $1/\Gamma$ (shown in gray) within the jet (shown in purple). The polarization vectors of the photons are indicated as black arrows with the direction indicating the PA and the size representing the PD. For an observer for whom the observer area is contained within the jet, the ordered magnetic field lines allow to observe a significant level of polarization. In case the observed area becomes equal to the size of the jet, the polarization will start to cancel out, resulting in a low PD.}
  \label{fig:tor_jet}
\end{figure}

The situation changes in case the magnetic field in the jet becomes ordered. Specifically, in the case of an ordered toroidal magnetic field, the polarization vectors are oriented around the magnetic field lines as illustrated in \autoref{fig:tor_jet}. In this scheme, which shows the magnetic field lines within the jet in purple, the PA is no longer symmetric around the viewing axis. As a result, a net polarization is achieved also when the observed area is fully contained in the jet. Such a magnetic field structure therefore predicts a non--zero PD when $\theta_v < \theta_c$ \citep{Gill2018}. For larger observation angles the PD drops, as the level of asymmetry is lost.

\subsection{Polarization Angle}

Although changes in the polarization angle throughout a GRB have been reported (see discussion in \autoref{sec:intro_PDP}) and could give insights into the emission model \citep{Gill:2021jzc}, when observing a constant PA throughout a GRB this parameter has no discriminative power. This is due to the fact that, without a reference, the PA cannot be measured relatively to a known axis. However, this is not the case when combining with GW measurements since, in two of the major emission models, photospheric and synchrotron, the PA will be related to the polarization angle $\psi$. We will provide a qualitative description of the PA for both models below.

\subsubsection{Photospheric emission}

\begin{figure}
  \centering
  \includegraphics[width=0.4\linewidth]{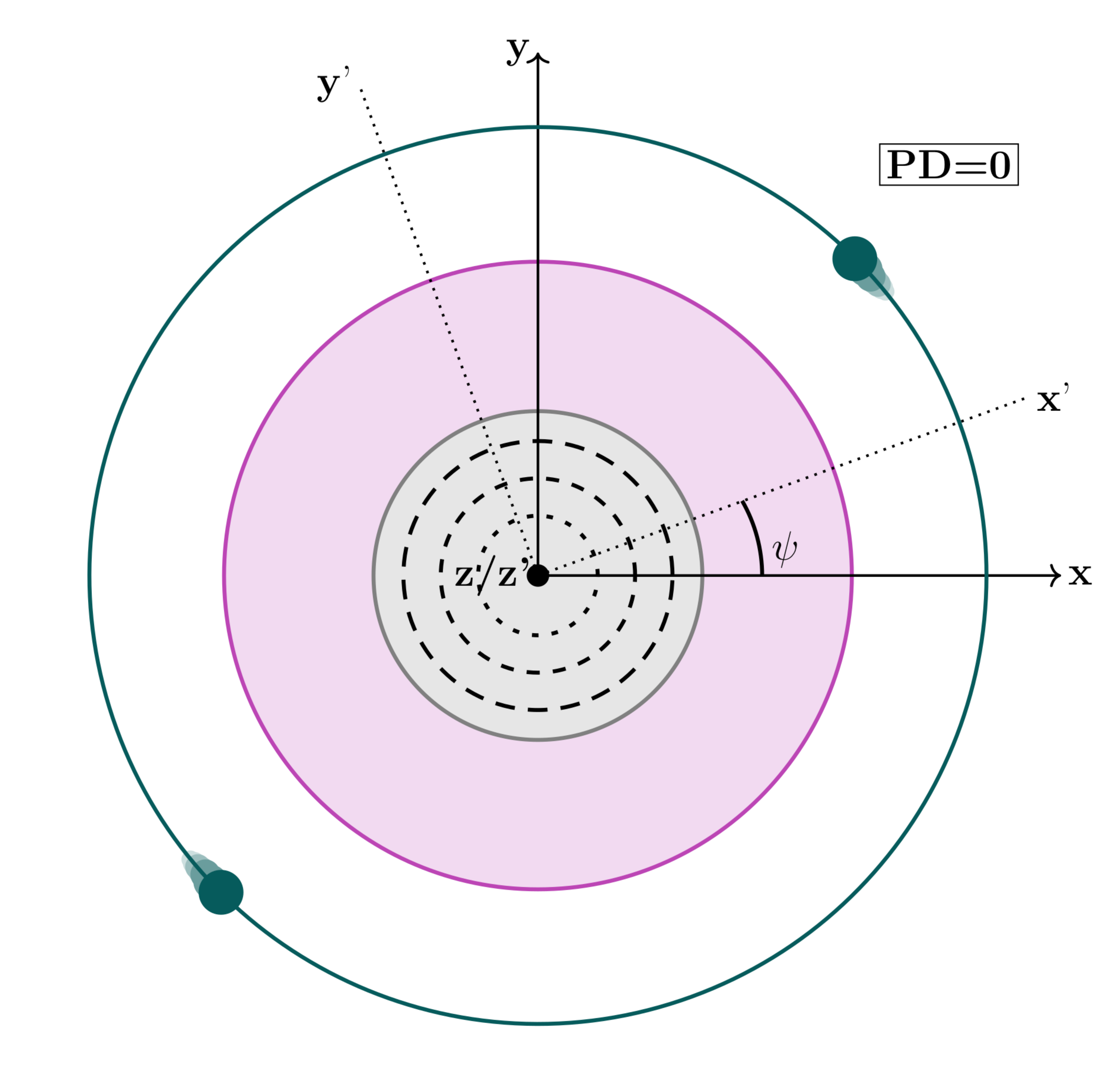}
  \caption{Illustration of the photospheric emission when observing the GRB with $\iota=\SI{0}{\degree}$. The observed area with a size of $1/\Gamma$ is shown in gray and is fully contained within the jet. As discussed before, we do not see any PD from this angle and therefore the PA is undefined.}
  \label{fig:on-axis}
\end{figure}

\begin{figure}
  \centering
  \includegraphics[width=0.7\linewidth]{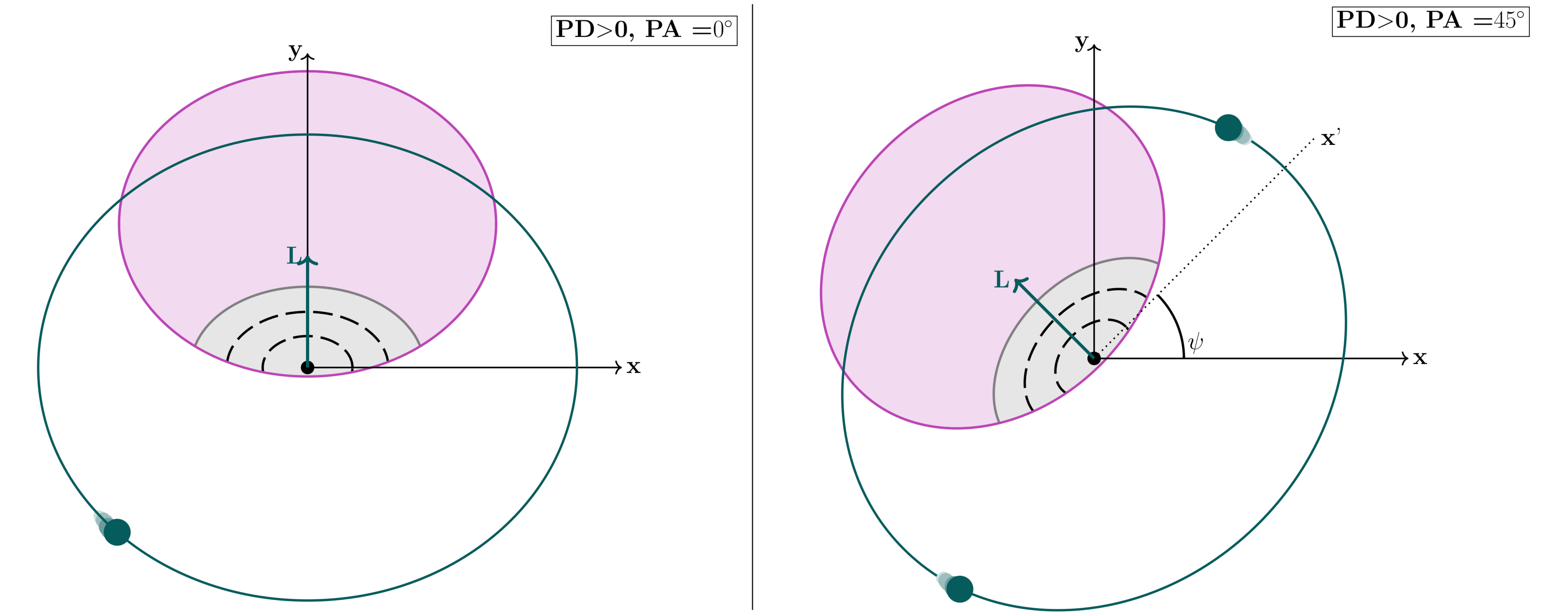}
  \caption{\textbf{\emph{Left}}: Illustration of the photospheric emission when observing the GRB with $\iota>\SI{0}{\degree}$ and $\psi=\SI{0}{\degree}$. The observed area with a size of $1/\Gamma$ is in gray and is no longer fully contained in the beam allowing for PD>0. The polarization of the remaining emission areas no longer cancels out, and the PA is aligned with $\psi$. \textbf{\emph{Right:}} The same as left but now with  $\psi=\SI{45}{\degree}$ resulting in a PA of $\SI{45}{\degree}$. }
  \label{fig:off-axis_psi0}
\end{figure}

For small $\iota$ no polarization is observed, and therefore the PA is not defined. This is illustrated in \autoref{fig:on-axis}, which shows the orbital plane of the progenitor system in green, the jet pointing towards the observer in purple and the observed area of size $\Gamma^{-1}$ in gray. The polarization vectors within the observed area are illustrated here with black lines, where the PA is in the direction of the lines and the PD by the size of these arrows. In this image the polarization angle $\psi$ is illustrated, however, since $\iota=\SI{0}{\degree}$, the PA is undefined. This changes when $\iota$ becomes such that the rotational symmetry of the observed area is lost, as illustrated on the left in \autoref{fig:off-axis_psi0}. In this image $\iota$ is chosen so that $\theta_v > \theta_c$, resulting in a net PD. The remaining polarization vectors inside of the observed area no longer cancel out, and a net PD with a PA along the $x$-axis is observed. In this case the PA therefore aligns with $\psi$. Finally, we can consider a rotated system, for example, setting $\psi=\SI{45}{\degree}$ as shown on the right in \autoref{fig:off-axis_psi0}. Again the symmetry is broken, however, in this scenario the PA will be $\SI{45}{\degree}$. We can therefore see that, for photospheric emission, the PA is aligned with $\psi$.

We can conclude that generally, in case a non--zero PD is observed, photospheric emission would produce a PA equal to $\psi$. A small note should however be added: for $\iota$ slightly below $\theta_c$, a small net PD will be observed as a small part of the $\Gamma^{-1}$ will be out of the jet. For these angles the remaining PA will be equal to $\psi + \pi/2$. This is however the case only for a limited range of $\iota$ and the PD observed will be very low \citep{Gill2018}. 

\subsubsection{Synchrotron}

For synchrotron emission from a random magnetic field the situation is similar to the photospheric case. Again, as the PD is zero for $\iota < \theta_c$, the PA is not defined in this region. Only when $\iota > \theta_c$ the PD will be non--zero and, similar to what was shown for photospheric emission, a PA is observed due to cancellation of part of the symmetry in the emission region. Contrary to photospheric emission however, $\mathrm{PA}=\psi + \pi/2$, allowing for a clear distinction between the two emission models. Also here, it should be noted that, for $\iota$ approaching $\theta_c$, a small PD will be observed with $\mathrm{PA}=\psi$. 

Synchrotron emission from a toroidal magnetic field produces a non--zero PD when $\iota < \theta_c$. As can be observed in \autoref{fig:tor_jet}, here $\mathrm{PA}\sim\psi + \pi/2$ and therefore almost equal to that for synchrotron from a random magnetic field. However, the two models can be distinguished from one another since the emission from a toroidal field will only produce a non--zero PD for $\iota < \theta_c$, while in case of a random magnetic field the opposite is true. It should also be noted here that the exact polarization angle for a toroidal magnetic field is not exactly  $=\psi + \pi/2$, and its deviation from this depends on several details. The exact angle is given by \citep{Granot:2004ys,Gill2018}:
\begin{equation}
    {\rm PA} = \phi_\Gamma - \arctan{\left[\frac{1-\xi}{1+\xi}\left(\frac{\sin\phi_\Gamma}{a+\cos\phi_\Gamma}\right)\right]}\,,
\end{equation}
where, in the chosen coordinate system, $\xi=(\Gamma\ \theta_e)^2$ and $a=\theta_e/\iota$ with $\theta_e$ being the angle between the local bulk velocity in the jet and the observed photons, and $\phi_\Gamma$ is the second angle defining the bulk velocity vector in the jet frame. For our qualitative discussion here $\phi_\Gamma\approx\psi+\pi/2$, and the PA can be trivially shown to be centered around $\psi+\pi/2$.

Apart from emission from a random magnetic field and from a toroidal field, various other magnetic field configurations can be considered, such as the emission from an ordered magnetic field within the jet plane ($B_{\parallel}$). Whereas the PD for such emission shows similar characteristics to that from synchrotron emission from a random field, except with a higher PD \citep{Granot:2003hx, Granot:2003dy, Gill2018}, the PA here is fully defined by the magnetic field direction. As this direction is very likely not correlated to the progenitor system in case of a BNS, the PA and $\psi$ will then show no clear correlation in this scenario.

\section{Future Prospects}\label{sec:future_prospects}

\begin{figure}
  \centering
  \includegraphics[width=14cm]{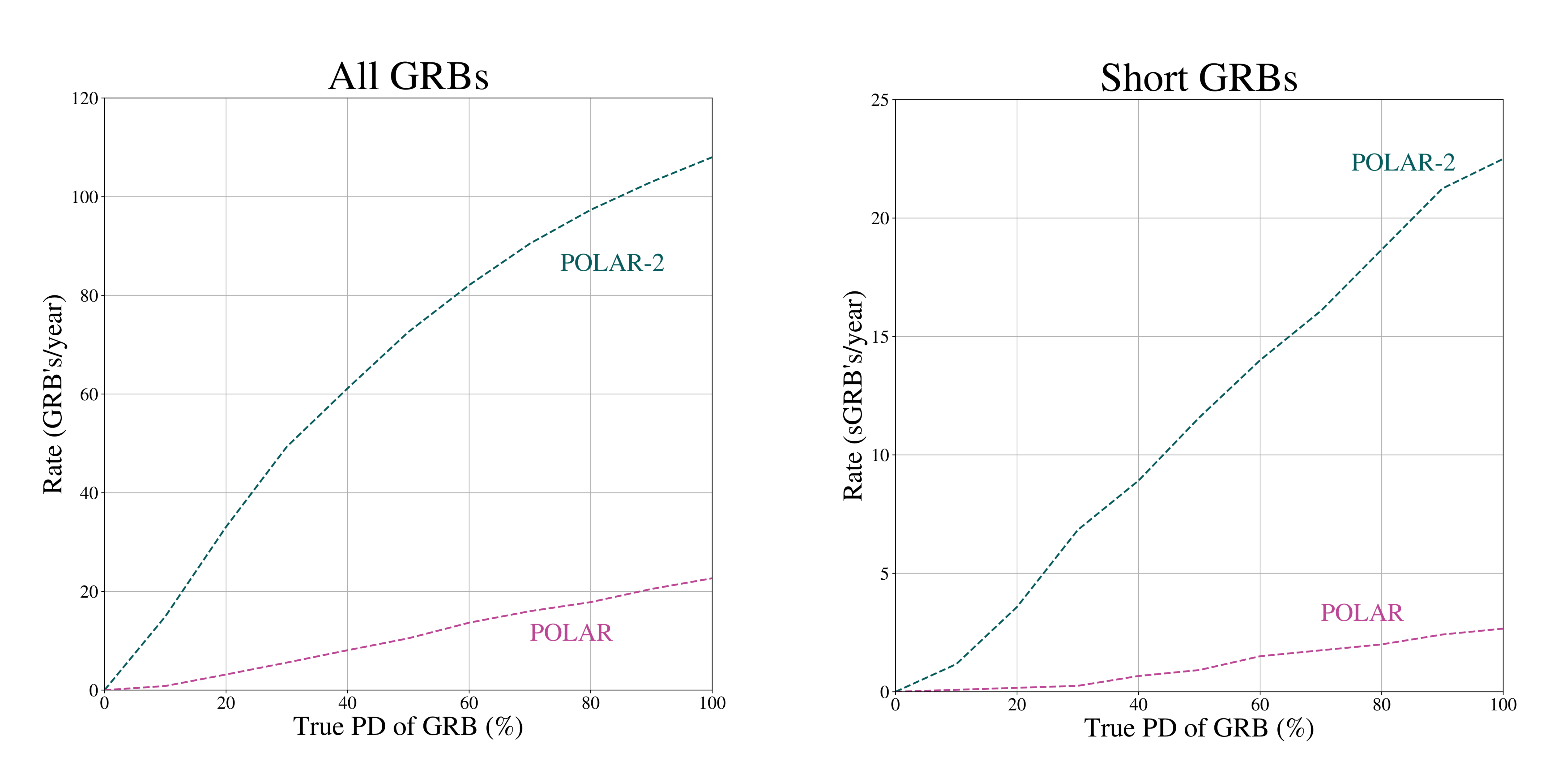}
  \caption{The number of total GRBs (left) and short GRBs (right) for which POLAR-2 will be able to distinguish the emission from being non--polarized as a function of their true PD. The predictions for POLAR calculated using the same method are also shown. The POLAR predictions for all GRBs match the measured results well, while for short GRBs the predictions are slightly lower than what was measured (POLAR got 4 measurements with an MDP<$70\%$).}
  \label{fig:MDP_POLAR-2}
\end{figure}

\begin{figure}
  \centering
  \includegraphics[width=7cm]{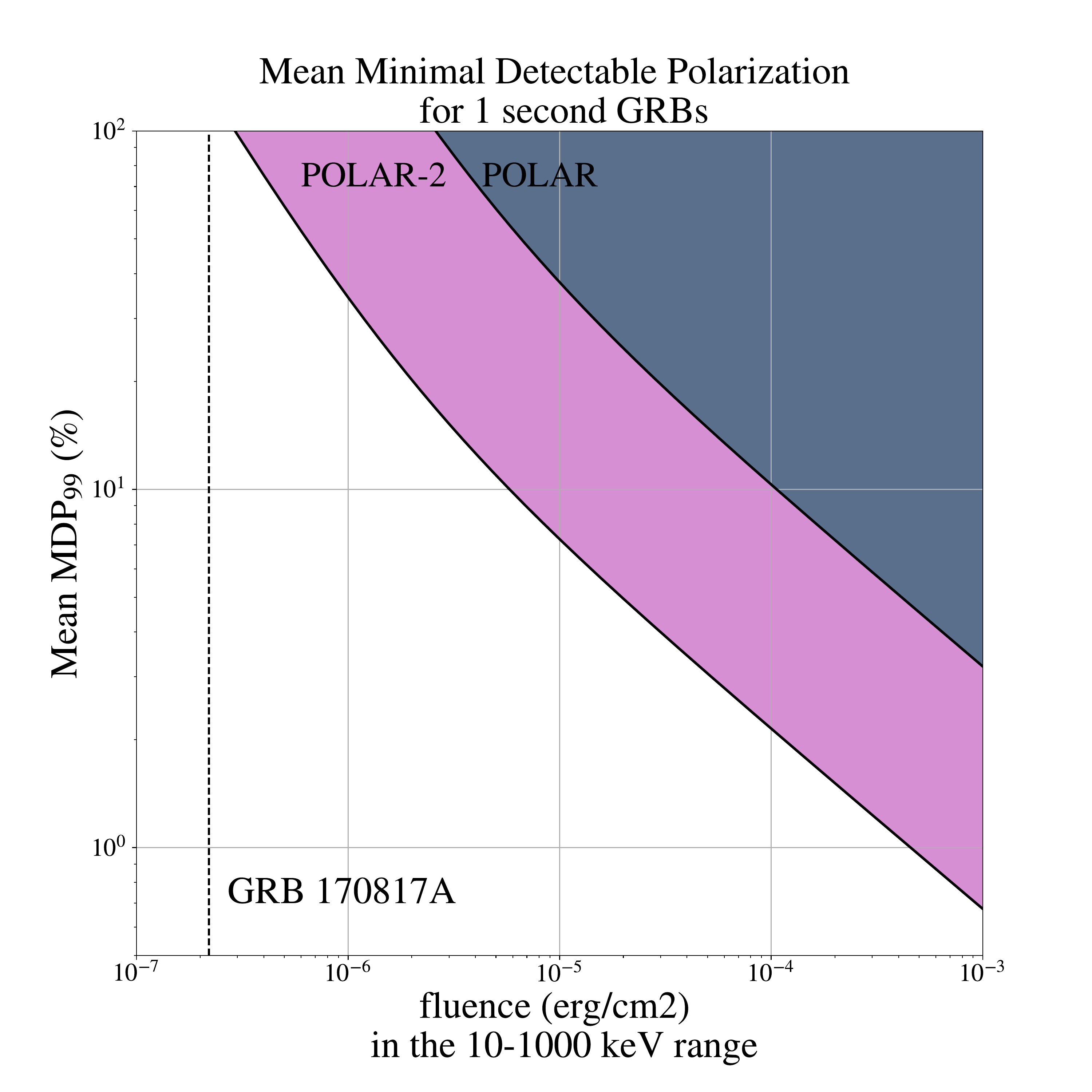}
  \caption{The Mean MDP(99\%) (the MDP averaged over the polarization angle) for POLAR-2 and POLAR for 1 second long GRBs as a function of the fluence of the GRB. The fluence of GRB 170817A is added for illustrative purposes.}
  \label{fig:sens_POLAR-2}
\end{figure}

\subsection{Gamma--ray Polarimetery for BNS}

To date, two dedicated instruments for studying the polarization of the prompt emission of GRBs have been launched, GAP \citep{GAP1} and POLAR \citep{POLAR1}. While the former reported polarization measurements for a handful of measurements \citep{GAP2,GAP3}, the latter detected a total of 55 GRBs during a 6 months period, and was able to perform polarization measurements for 14 of these \citep{Kole:2020eef}. In addition the AstroSAT CZT instrument, which was not designed as a polarimeter but was shown on ground to be capable of polarimetry, was used to produce a total of 20 GRB polarization measurements throughout approximately 5 years of data taking \citep{Chattopadhyay:2022yzx}. The sensitivity of all these instruments is limited, with the most sensitive of the three being POLAR, which had an effective area  of approximately \SI{400}{\square\centi\meter} at \SI{300}{\kilo\eV} for polarization measurements and \SI{500}{\square\centi\meter} for spectrometry. 

Out of the 14 polarization measurements reported by POLAR, 4 were classified as short GRBs, and can therefore likely be associated to BNS mergers \citep{Kole:2020eef}. The 4 GRBs were 170101A, 170127C, 170206A and 170305A of which the brightest, 170206A, had a fluence of \SI{1e-5}{\erg\per\square\centi\meter} and allowed to produce a measurement of PD of $13.5\substack{+7.5 \\ -8.6}\%$ whereas 170305A was an order of magnitude weaker and only allowed to provide an upper limit for the PD of around $65\%$. It is therefore clear that for GRBs such as 170817A, which was almost another order of magnitude weaker with a fluence of \SI{2e-7}{\erg\per\square\centi\meter}, no constraining polarization measurement would have been possible. It should be noted that, based on its sensitivity, POLAR would have had a significant detection of this GRB, but the number of detected photons would have been insufficient for polarization measurements. For this purpose, more sensitive gamma--ray polarimeters are being developed in the form of POLAR-2 and LEAP. 

Although the design of these instruments is different, with LEAP using 2 different types of scintillators whereas POLAR-2 only uses 1, their effective areas are rather similar, with LEAP having an effective area (for polarization events) of around $1000\,\mathrm{cm^2}$ and POLAR-2 of $1200\,\mathrm{cm^2}$ for polarization events, and about twice as much for spectrometry \citep{Kole:2022gxs}. Furthermore the energy range of POLAR-2 is extended down to \SI{20}{\keV}, whereas for POLAR the low energy limit was \SI{50}{\keV}. The capability to measure polarization is generally quantified using the Minimal Detectable Polarization (MDP) \citep{Weisskopf:2010se}, defined as:
\begin{equation}
    \mathrm{MDP} = \frac{2\sqrt{\mathrm{-ln}(1-{\rm C.L.})}}{M_{100}C_s}\sqrt{C_s+C_b}\,.
\end{equation}

where $C_s$ is the number of signal counts, $C_b$ is the number of background counts, $M_{100}$ is an instrument dependent term, quantifying the relative modulation measured by the detector for a $100\%$ polarized flux and C.L. is the requested confidence level.\footnote{For wide field of view instruments like POLAR-2 and LEAP the MDP definition has some issues, as outlined in \cite{Kole:2022gxs}; however, for our discussion here these are not relevant.} The MDP provides the minimum level of polarization one can exclude with a confidence level (C.L.) from being non--zero using the detector. Using the simulated instrument performance of POLAR-2, as well as the GRB catalog of \textit{Fermi}-GBM, we can calculate the number of GRBs POLAR-2 would see per year with a certain MDP. This is shown in the left panel of \autoref{fig:MDP_POLAR-2}, along with that of POLAR, calculated in the same way for comparison. The numbers for POLAR match with the performance observed from the actual POLAR data, thereby indicating this method to be reliable. Of course, it should be noted that for POLAR-2 the performance parameters, outlined e.g. in \cite{POLAR-2:2021uea}, have not yet been instrumentally verified, and are based on simulations only. We can also restrict the number of measurements to only those for short GRBs; the result in this case is shown in the right panel of \autoref{fig:MDP_POLAR-2}. We can see that POLAR-2 will be able to perform rudimentary polarization measurements for approximately 20 short GRBs per year, while it will be able to provide detailed measurements for several events per year. This is further indicated in \autoref{fig:sens_POLAR-2} which shows the MDP (with $\mathrm{C.L.=99\%}$) of POLAR-2, as well as POLAR for a verifiable comparison, as a function of the fluence for 1 second long GRBs. The fluence of 170817A is highlighted to indicate that POLAR-2 will be capable of performing rudimentary polarization measurements for short GRBs slightly brighter than 170817A.

Although for LEAP sufficient details to perform such an analysis are not available, we expect a similar performance to POLAR-2. Due to the instrument design, the field of view of LEAP is smaller than that of POLAR-2 (as the high--Z scintillators block photons with large incoming angles with respect to the instrument pointing axis), resulting in an overall lower rate. However, the inclusion of the high--Z scintillators do increase the $M_{100}$, thereby allowing for more precise measurements for the GRBs which are observed. Overall, we therefore expect a slightly lower MDP than that of POLAR-2 presented in \autoref{fig:MDP_POLAR-2}, whereas the number of observations will be lower.

\subsection{Future GW observation possibilities}\label{sec:future_GW}

In this section, we forecast the capabilities of measuring the polarization angle $\psi$ and the inclination $\iota$ during the upcoming observing runs of the present network of GW observatories, as well as with third--generation ground--based instruments. 

We simulate a catalog of BNS sources as follows: the sources are uniformly distributed in the sky, the source--frame masses of the two objects are drawn independently from a flat distribution between \SI{1}{\Msun} and \SI{2.5}{\Msun}, the aligned spin components are drawn from two independent flat distributions in the range $[-0.05, 0.05]$, and the redshift distribution is given by a Madau--Dickinson profile \citep{Madau:2016jbv}, convolved with a time delay $P(t_d)\propto1/t_d$, and with the local rate fixed to the median value obtained combining the observations of the first three observing runs of LIGO, Virgo and KAGRA, ${\cal R}_{0, {\rm BNS}}=\SI{105.5}{\per\cubic\giga\parsec\per\year}$ \citep{LIGOScientific:2021psn}. The details and the values obtained for the parameters of the redshift distribution are given in Appendix A.1 of \cite{Iacovelli:2022bbs}.
With these choices, the total number of mergers expected in one year up to $z\sim20$ is about \num{e5}, which is the number of sources we generate.

To obtain forecasts for the the measurement of the signal parameters for this population of sources by a network of GW detectors, we use the Fisher--matrix based parameter estimation code \texttt{GWFAST} \citep{Iacovelli:2022mbg}.\footnote{Other publicly available softwares that implement the Fisher matrix formalism for GW detector networks are \texttt{GWBENCH} \citep{Borhanian:2020ypi} and \texttt{GWFISH} \citep{Harms:2022ymm} [see also \cite{Chan:2018csa,Pieroni:2022bbh}]; all these have been shown to be in good agreement among them and with \texttt{GWFAST} \citep{Iacovelli:2022bbs}.}
The Fisher analysis is performed only for events having a network ${\rm SNR}\geq12$. We further take into account the effect of imposing realistic priors on the GW parameters, which is particularly relevant for events with $\iota\sim0$ or $\iota\sim\pi$ which are those more interesting for this paper (see the discussion in \cite{Iacovelli:2022bbs}). The details of the analysis are given in \autoref{sec:appendix_fisher}.

 For second generation GW detectors, results are presented first for the O5 run of the current network of detectors, i.e. the two LIGO detectors in the U.S. the Virgo detector in Italy, and the KAGRA detector in Japan, together with LIGO India\footnote{The sensitivity curves are available at \url{https://dcc.ligo.org/LIGO-T2000012/public}, \cite{AbbottLivingRevGWobs}.} foreseen to start in 2026; then, we consider the same network with upgraded LIGO detectors in the Voyager design\footnote{The sensitivity curves are available at \url{https://dcc.ligo.org/LIGO-T1500293/public}.} foreseen towards the end of this decade. We also report the predictions for the next generation of ground--based detectors. The prediction for the Einstein Telescope (ET),\footnote{The ET--D sensitivity curve is available at \url{https://apps.et-gw.eu/tds/?content=3&r=14065}. See also \url{https://www.et-gw.eu/index.php/observational-science-board} for a repository of papers relevant for ET, produced in the context of the Observational Science Board (OSB) activities.} which is foreseen to become operational in the next decade, as well as the predictions for a combined run of ET and the two future Cosmic Explorer (CE) detectors are shown (in particular, we consider the CE network to be made up of a \SI{40}{\kilo\meter} and a \SI{20}{\kilo\meter} instrument).\footnote{The sensitivity curves are available at \url{https://dcc.cosmicexplorer.org/CE-T2000017/public}}. The two are located at the LIGO Hanford and LIGO Livingston sites respectively. For the O5 and Voyager configurations we assume the 5 detectors to be operational 70\% of the time, while for 3G detectors we assume a 85\% duty cycle.
\begin{figure}[t]
    \centering
    \begin{tabular}{c@{\hskip -3mm}c}
         \includegraphics[width=70mm]{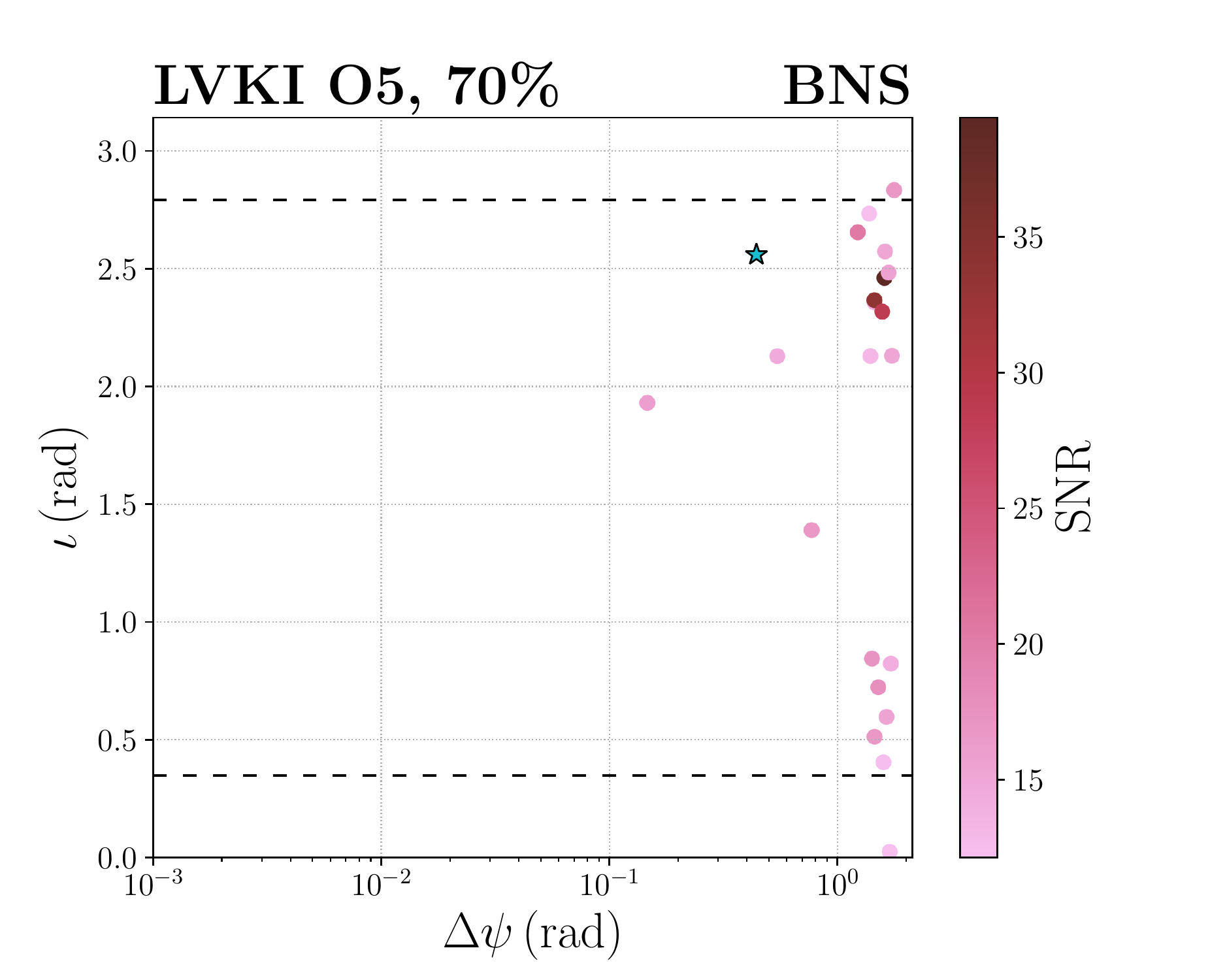} & \includegraphics[width=70mm]{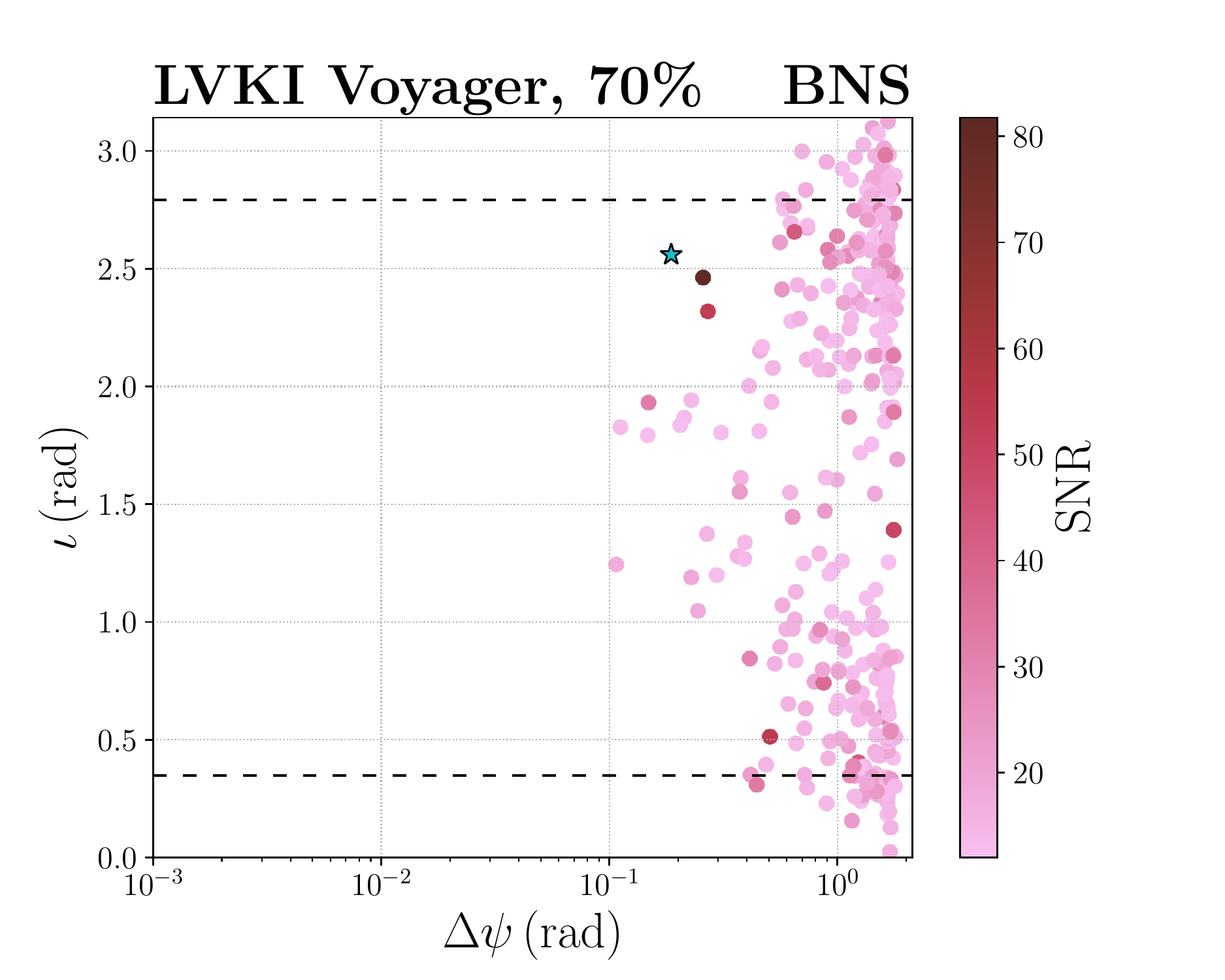} \\
         \includegraphics[width=70mm]{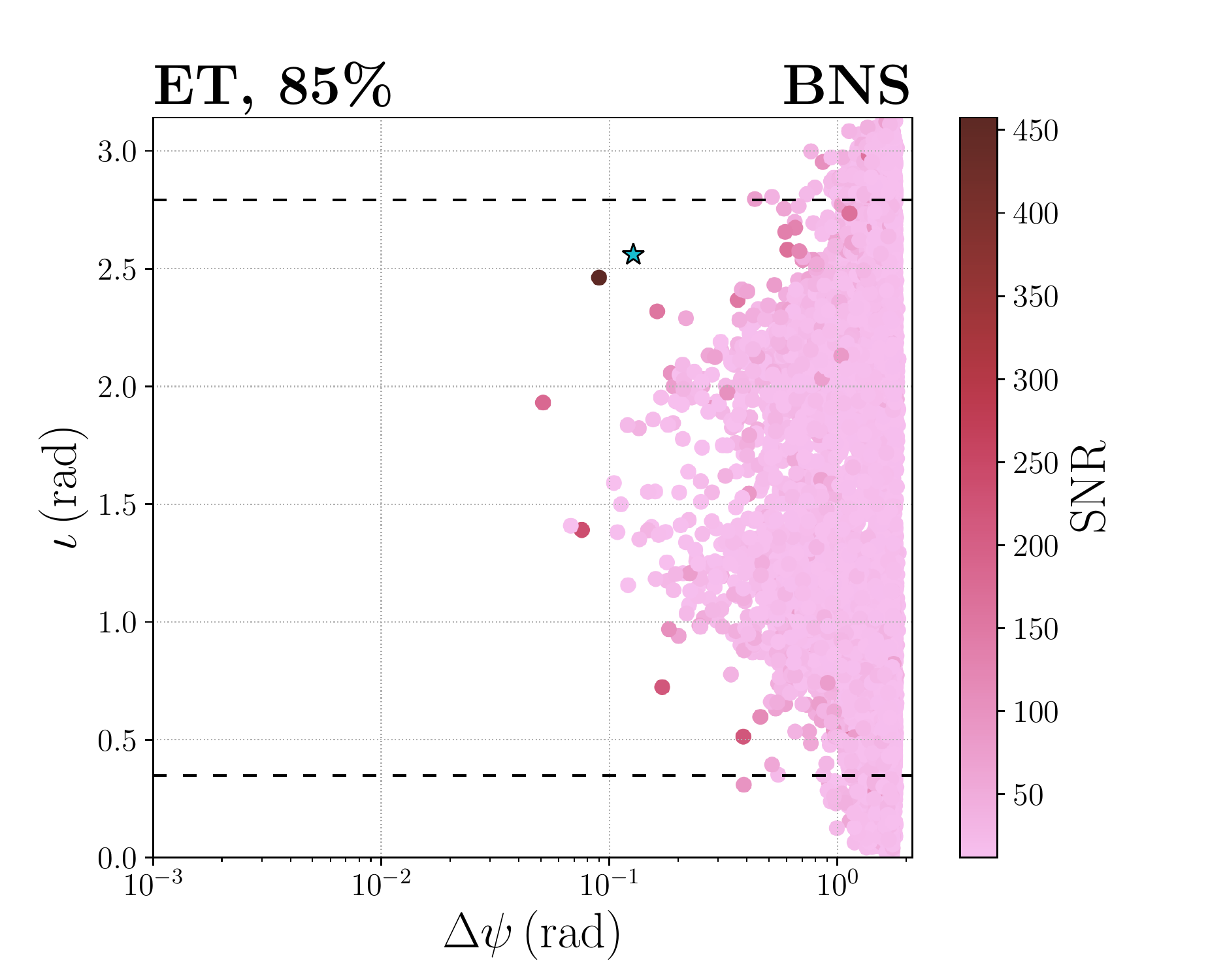}  & \includegraphics[width=70mm]{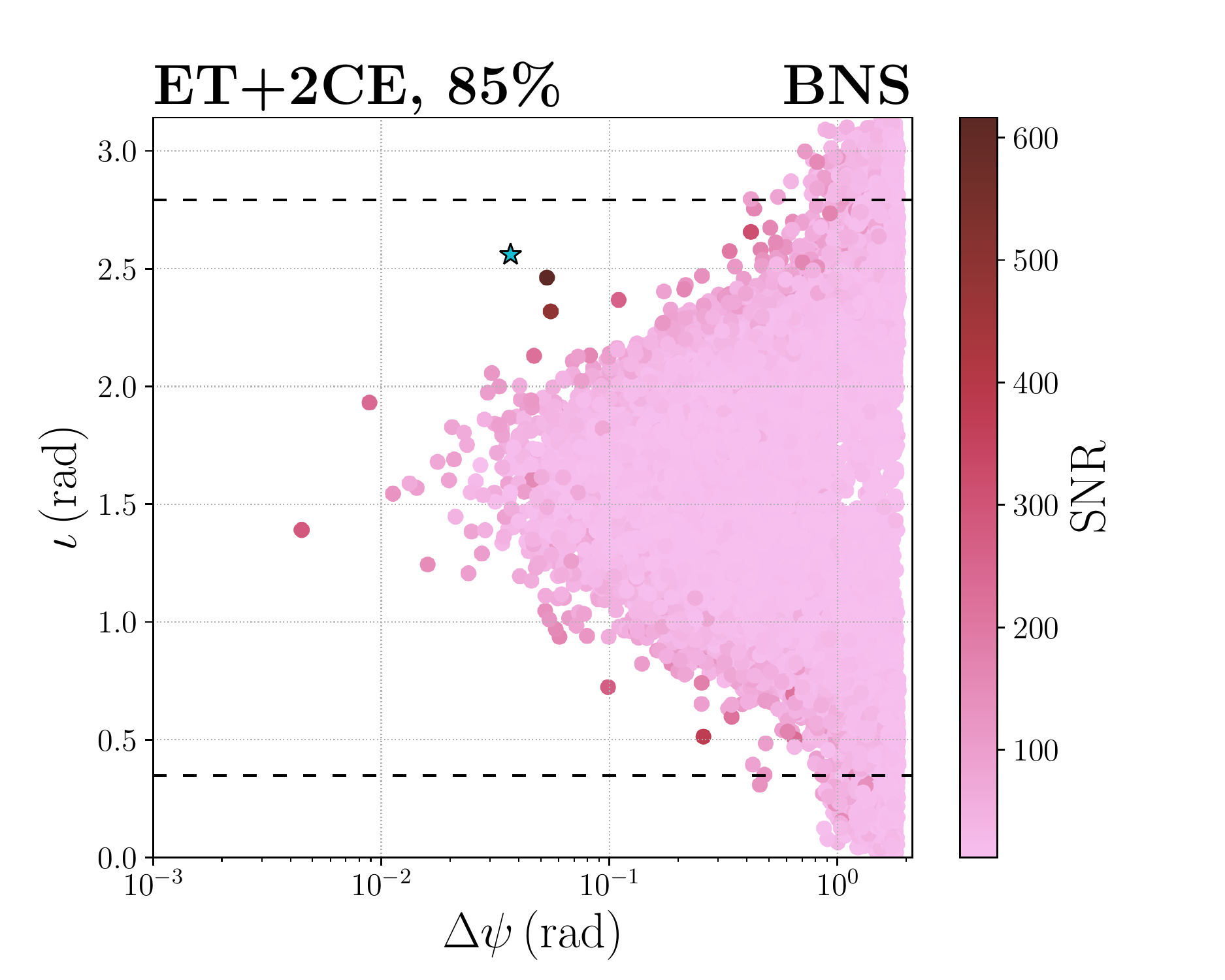}
    \end{tabular}
    \caption{Scatter plots of the observed BNS population in one year at various detector networks. On the first row we report the results for the current network of detectors, namely LIGO, Virgo and KAGRA, plus LIGO India both during the O5 run (left panel) and in the Voyager stage for the LIGO detectors (right panel). In the bottom row we instead report the results for 3G detectors, both for ET alone (left panel) and a network consisting of ET and two CE detectors (right panel). In each panel we report the observed errors on the polarization angle $\psi$ as a function of the inclination angle $\iota$, with a color code that gives information on the signal--to--noise ratio. The dashed lines in each panel mark $\iota=\SI{20}{\degree}$ and $\iota=\SI{160}{\degree}$, to give an hint of the regions where a beamed counterpart is observable. The percentages in the titles refer to the assumed (independent) duty cycle of each detector in the considered network configuration. For reference, we report the estimated values for the GW170817--like event described in the text as a blue star.}
    \label{fig:3G_prospects_psi}
\end{figure}

\begin{figure}[t]
    \centering
    \begin{tabular}{c@{\hskip -3mm}c}
         \includegraphics[width=70mm]{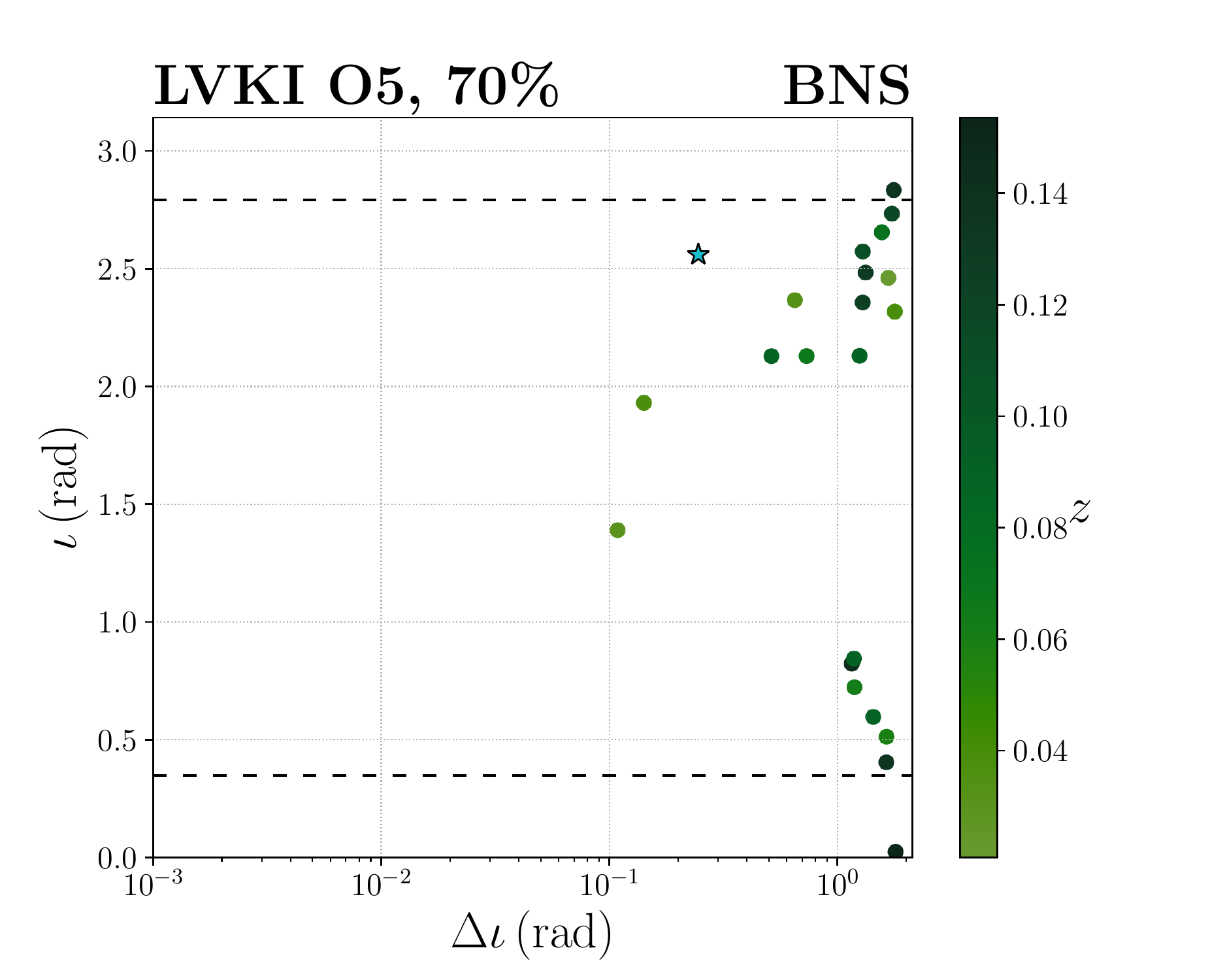} & \includegraphics[width=70mm]{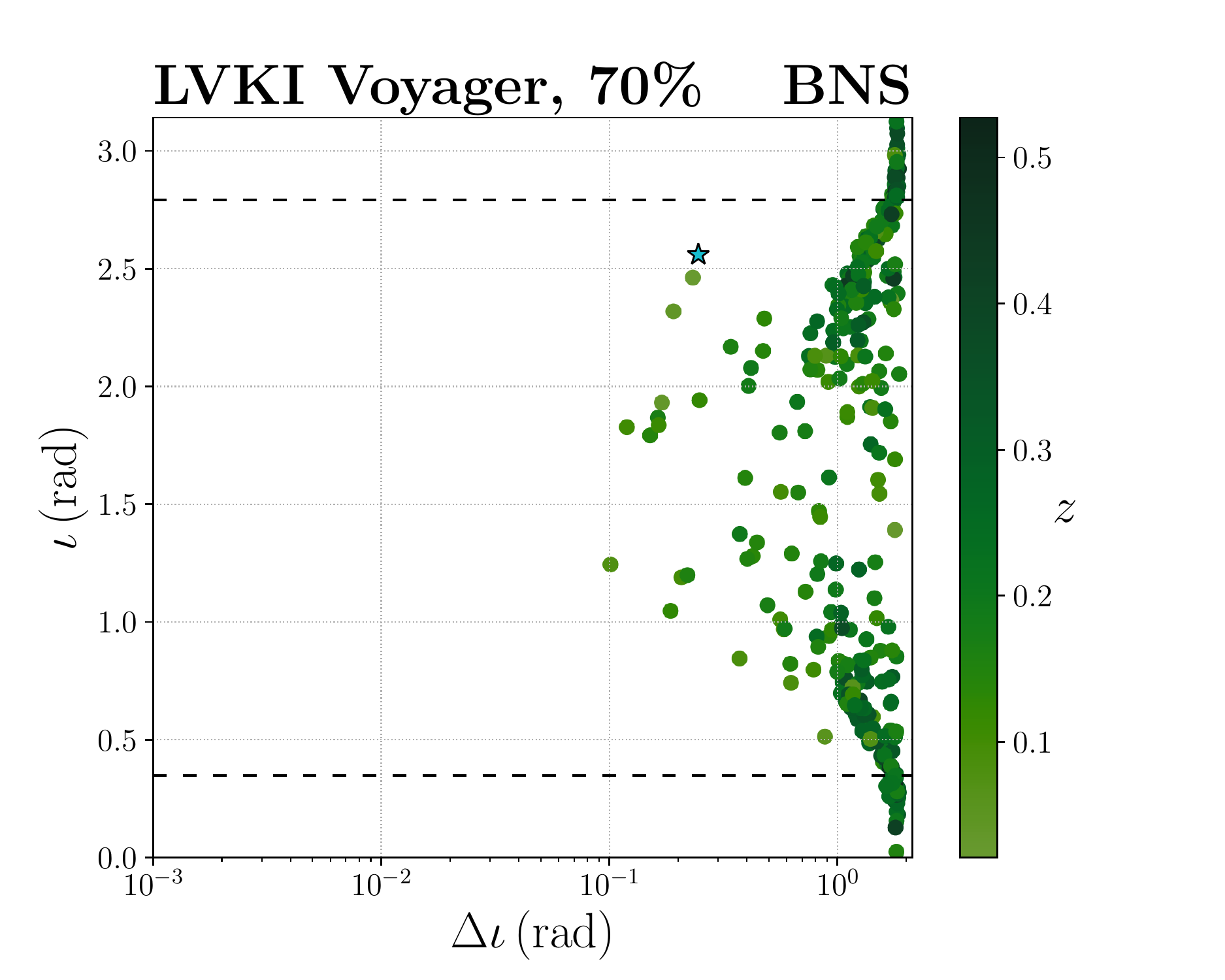} \\
         \includegraphics[width=70mm]{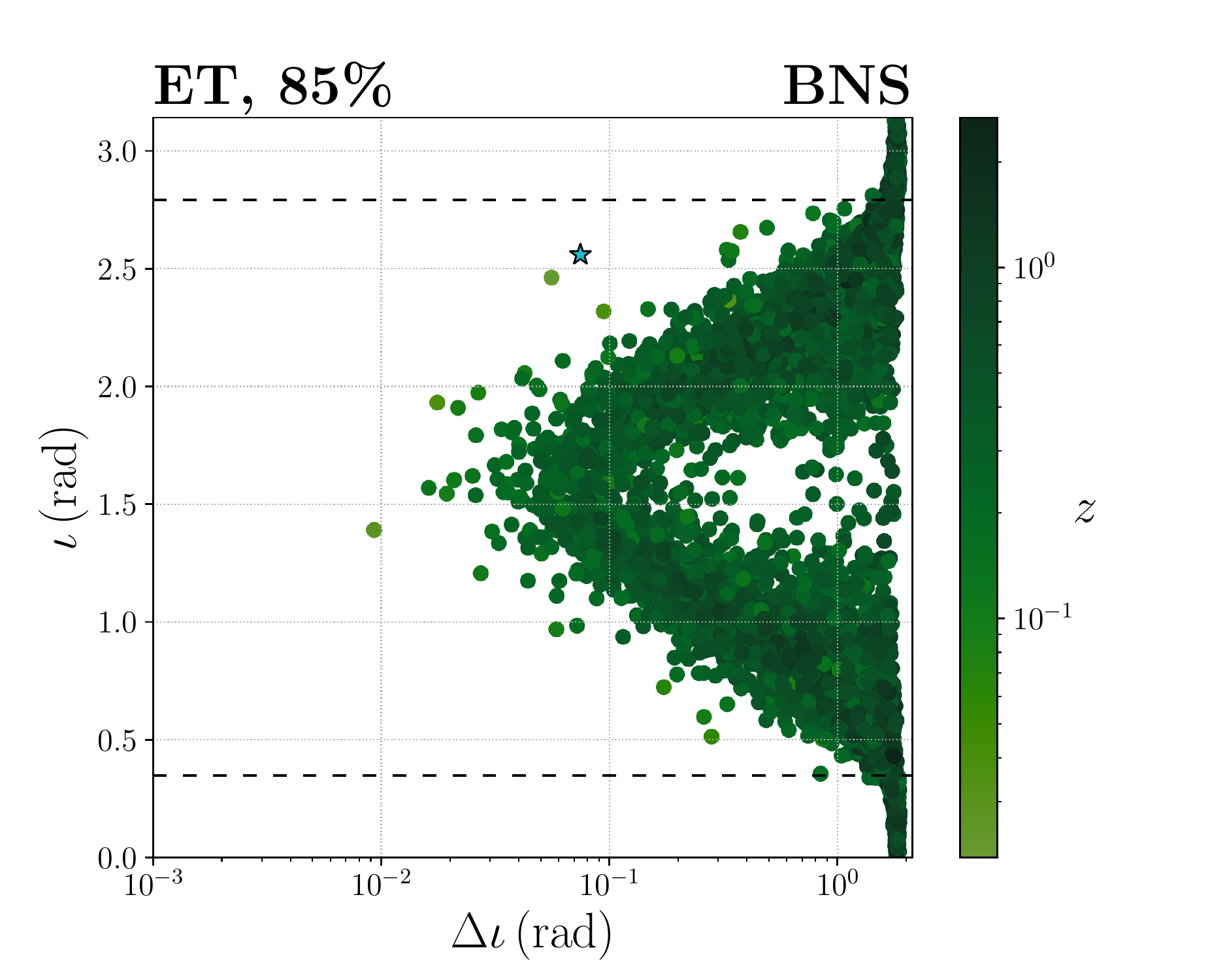}  & \includegraphics[width=70mm]{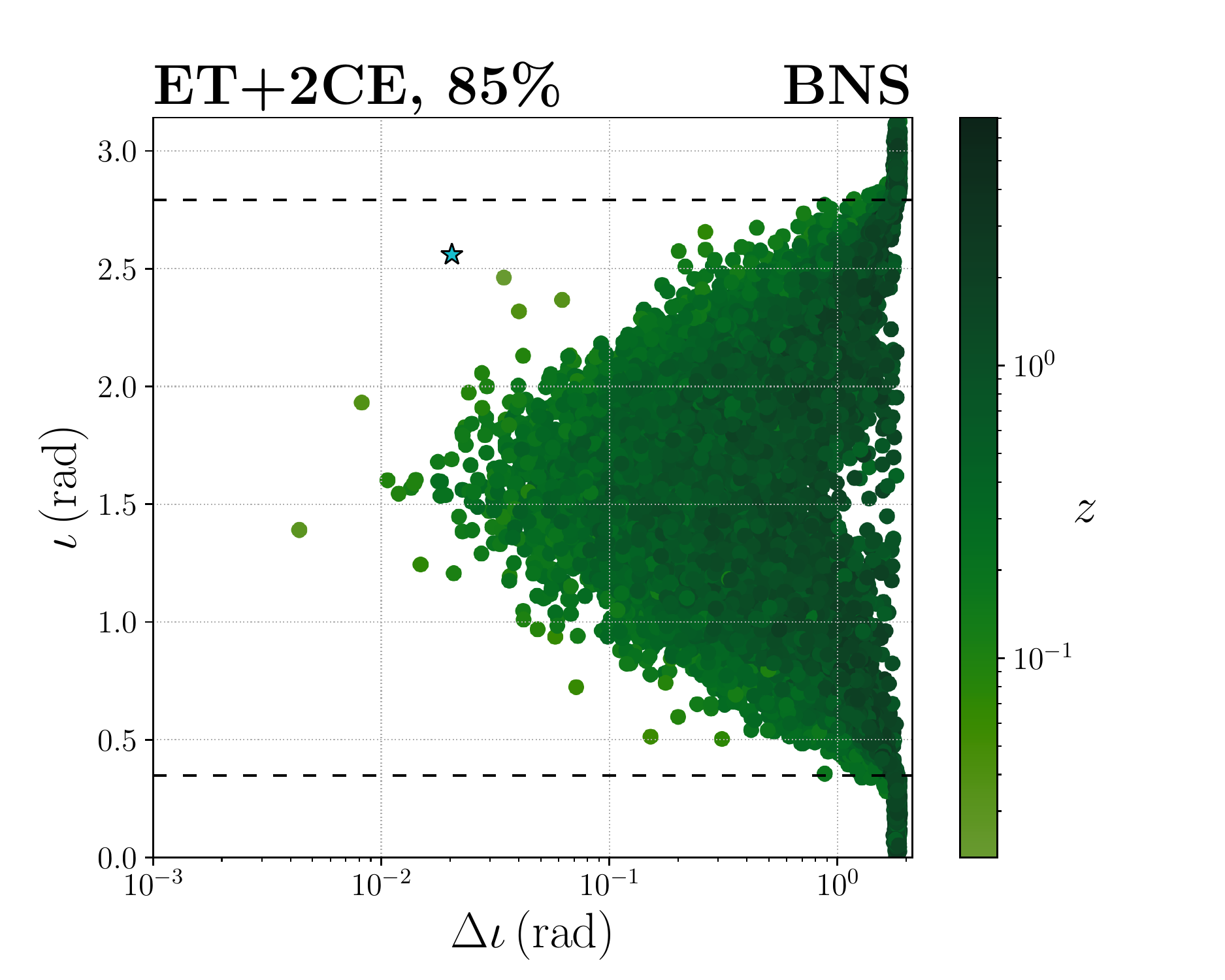}
    \end{tabular}
    \caption{As in \autoref{fig:3G_prospects_psi} for the observed errors on the inclination angle $\iota$ as a function of $\iota$, with a color code that gives information on the redshift of the binary.}
    \label{fig:3G_prospects_iota}
\end{figure}

In \autoref{fig:3G_prospects_psi} we show the forecasted $1\sigma$ uncertainty $\Delta \psi$ attainable by these four different detector networks on the parameter $\psi$, correlated with the value of the inclination angle $\iota$ on the vertical axes. The color bar represent the SNR of the detection. We see that, for second--generation detectors, the uncertainty on $\psi$ is largely prior--dominated, as most of the forecasted values of $\Delta \psi$ align with the maximum value allowed by the prior range (taken uniform in $[0, \pi]$), corresponding to the upper limit of the $x$ axes in the plots, and only a handful of events have errors going down to about \SI{10}{\degree}. For third--generation detectors instead measurements with $\Delta \psi\lesssim \SI{5}{\degree}$ accuracy would be possible for a large number of sources, with a network made by the ET and two CE. 
\autoref{fig:3G_prospects_iota} shows the same results for the forecasted $1\sigma$ uncertainty on the inclination angle, $\Delta \iota$. In this case, similar considerations as for $\Delta \psi$ apply for 2G detectors, while for 3G detectors it is worth noticing that competitive measurements of $\iota$ ($\Delta\iota\lesssim \SI{10}{\degree}$) will be possible for a large number of sources: about 600~ev/yr even with a single ET detector, and about 3000~ev/yr with the network ET+2CE. 

It is interesting to compare these results on a full population with the best available actual BNS detection to date, i.e. GW170817. Hence, we perform the Fisher analysis for a GW event with parameters equal to the median ones estimated for GW170817, \footnote{The posterior samples are available on the GWOSC website, \url{https://www.gw-openscience.org/eventapi/html/GWTC-1-confident/GW170817/v3}.} namely: $m_1=\SI{1.51}{\Msun}$, $m_2=\SI{1.25}{\Msun}$, $\theta=\SI{113}{\degree}$, $\phi=\SI{197}{\degree}$, $d_L=\SI{43}{\mega\parsec}$, $\iota=\SI{147}{\degree}$, $\psi=\SI{45}{\degree}$ (this value of $\psi$ is actually chosen for convenience, since this parameter is unconstrained). With these choices, we estimate that such a signal could lead to a measurement of $\iota$ with an error of about $\SI{12}{\degree}$ and of $\psi$ of about $\SI{25}{\degree}$ at the 2G network during O5 (the network SNR of such an event is about 158 with this detector configuration), while using the LIGO detectors at the Voyager stage we find $\Delta\iota\simeq\SI{7}{\degree}$ and $\Delta\psi\simeq\SI{11}{\degree}$, with a SNR of 374. For 3G detectors the accuracy is even higher: we estimate that, for a signal like GW170817, at ET alone it could be possible to obtain $\Delta\iota\simeq \SI{4}{\degree}$ and $\Delta\psi\sim \SI{7}{\degree}$ with a SNR of 644, which shows that this instrument alone can perform better than a network of five 2G interferometers. Finally, considering the ET+2CE network, we find that it would be possible to get $\Delta\iota\simeq \SI{1}{\degree}$ and $\Delta\psi\simeq \SI{2}{\degree}$ with a network SNR of 2261. If a polarization measurement of the counterpart of such a system was performed, it would have an outstanding potential capable of excluding the majority of the emission models thanks to its large $\iota$. In case the PD is high ($>50\%$), the only viable emission theory discussed here is synchrotron emission from a parallel magnetic field, while if the emission is partly polarized the emission can come either from Compton drag emission or synchrotron emission from a random magnetic field. Here the measurement of the PA allows to distinguish between these two ($\mathrm{PA}\sim\psi$ for Compton drag and $\mathrm{PA}\sim\psi+\pi/2$ for Synch. Rand.). Finally, for a low level of polarization the emission either has a photospheric origin or that from synchrotron emission from a toroidal magnetic field. Distinguishing between these two is possible using a detailed measurement of the PD, with photospheric emission allowing for higher PD (with $\mathrm{PA}=\psi$) than synchrotron emission.

The above forecasts suggest however that GW170817 lies in a tail of the distribution of possible sources, for which good measurements are possible.
This is also visible from \autoref{fig:3G_prospects_psi} and \autoref{fig:3G_prospects_iota}, where this event is shown as a blue star, and it lies outside of the bulk of the detected events (notice also that the reported SNRs for such an event are considerably higher than the ones characterising the events in the simulated population, reported in the color code of \autoref{fig:3G_prospects_psi}, mainly thanks to the small distance). 
It must thus be stressed that, according to current population models, events like GW170817--GRB170817A are not likely to happen often, and it could be difficult to observe another similar one in the coming years \citep{Colombo:2022zzp}. 

\subsubsection{Joint Detection Prospects}

While \autoref{fig:3G_prospects_psi} and \autoref{fig:3G_prospects_iota} show the estimated measurements for all BNS mergers observed by the various GW detector configurations, it will be possible to observe the EM counterpart only for a fraction of them, as the alignment between source and observer needs to be sufficiently favourable ($\iota\sim0$ or $\iota\sim\pi)$. The $\iota$ range in which an observation of the EM counterpart is possible is assumed here to be $\iota<\SI{20}{\degree}$ or $\iota>\SI{160}{\degree}$. These values are indicated by the dotted lines in \autoref{fig:3G_prospects_psi} and \autoref{fig:3G_prospects_iota}. The limit used here for $\iota$ is a compromise based on typical half jet opening angles of $\theta_c\sim\SI{10}{\degree}$ based on \cite{Salafia:2022dkz} and \cite{Alicia} and the fact that GRB 170817A was still visible to \textit{Fermi}-GBM with $\iota\sim\SI{150}{\degree}$ while a jet opening angle of $34^\circ$ was also reported for GRB 050724A \citep{Alicia}. It should be noted here that a jet origin of the emission from 170817A is not confirmed and the emission could instead come from a cocoon, explaining the low luminosity. The EM counterpart of GW170817 should therefore be considered as an outlier, and one cannot assume to be able to observe the emission from sources at such off--axis angles for every GRB. Generally, as was done in \cite{Yu:2021nvx}, the energy $E(\iota)$ emitted by the GRB at an angle $\iota$ is described to be correlated to its on--axis equivalent isotropic energy $E_0$:
\begin{equation}
E(\iota)=E_0 \exp(-\frac{\iota^2}{2\theta^2_c}).
\end{equation}
However, it should however be noted that much uncertainty remains regarding the details of this correlation, as well as on the typical opening angle of jet. For example, when comparing the joint detection predictions in \cite{Saleem:2019eii}, \cite{Howell:20197e} for 2G detectors, those in \cite{Yu:2021nvx} for both 2G and 3G detectors, and in \cite{Ronchini:2022gwk} for 3G detectors, one can see significant differences as a result of the variations in the assumptions made. The predictions obtained in the work presented here, which are made using a simplified model, should therefore only be taken as an order--of--magnitude indication on the possibility of joint detections. 

Based on the results from \autoref{fig:3G_prospects_psi} we can see that, during the mission time of POLAR-2, which is foreseen to coincide with the O5 run, only a handful of BNS mergers detected by the GW network will have a value of $\iota$ which would make them visible to gamma--ray detectors. Using the currently predicted performance of POLAR-2 and taking into account it observes only half of the sky, and only approximately $50\%$ of all BNS mergers is expected to produce a successful GRB (as found in \cite{Colombo:2022zzp}), we can derive that for POLAR-2 we expect around 0.25 to 2 joint detections per year with GW detectors. This number is compatible with the predictions for Fermi-GBM ($\sim0.4\,-\sim4$ joint detection per year) and GECAM ($\sim0.6\,-\sim5.6$ joint detection per year) presented in \cite{Yu:2021nvx}.
As POLAR-2 will have a larger effective area than Fermi-GBM and one equivalent to a single GECAM detector, while having a smaller field of view than these instruments, this prediction appears compatible. For LEAP a similar, although slightly lower, number of joint detections can be expected.
Once again we should however stress that these predictions rely on specific assumptions made regarding, to mention a few, the BNS populations, gamma--ray emission processes and jet opening angles. Providing detailed, model--independent predictions for the joint detection rate is not possible at the moment and the numbers provided here should only be taken as a rough estimation. The association between the GRB and the GW event is also required but, as it was the case for  GW170817--GRB170817A \citep{LIGOScientific:2017ync}, this is provided by a coincidence in the arrival times.\footnote{Notice, however, that the foreseen increase in the rate of GWs and short GRBs detections means that the false association probability based on time alone could increase. A significant association might therefore also benefit from a good localization for both the GW and the GRB, but we do not to include this aspect in the present analysis.}

It is clear from \autoref{fig:3G_prospects_iota} and \autoref{fig:3G_prospects_psi} that the values of $\iota$ and $\psi$ will be fully unconstrained for any joint detections during the O5 run (as the error on this parameter coincides with the full prior range), thereby not allowing to directly combine the GW and polarization measurements. 

Joint detections might become more numerous during the potential Voyager run. This is especially true if we assume the LEAP mission will be successfully launched in 2027 and the POLAR-2 mission will remain operational by this time, as this would provide a sky coverage for gamma--ray polarimeters of more than half the sky (the exact number depends on the relative orbit of the ISS and CSS). As can be seen from \autoref{fig:3G_prospects_psi} and \autoref{fig:3G_prospects_iota}, the number of joint detections could reach to several per year; however, constraining measurements of $\iota$ and $\psi$ remain out of reach for events with $\iota<\SI{20}{\degree}$ or $\iota>\SI{160}{\degree}$. Nonetheless, for far off--axis events such as 170817A, a single constraining measurement of both binary system angles could become possible as discussed in \autoref{sec:future_GW}.

Using instruments such as ET, the number of joint detections with polarimeters will significantly increase. While even here direct constraints from just the GW measurements of the $\iota$ and $\psi$ angles remain difficult for on--axis GRBs, the prospects for off--axis events such as GRB 170817A increase to around 10 joint detections per year for which GW measurements are capable of constraining $\iota$ and $\psi$ to several 10's of degrees. The detailed numbers, for both ET and other GW configurations are illustrated in \autoref{tab:summary_GW}. In order to deduce the possible joint detections with short GRBs these numbers should be corrected for the earlier mentioned probability of a BNS producing a successful GRB ($\sim50\%$) and for the field of view of the detector. Furthermore, EM follow--up measurements capable of localizing the host galaxy, and thereby allowing to further constrain $\iota$ and $\psi$, would greatly improve the situation. 

\begin{table}\label{tab:sum_GW}
\centering
\begin{tabular}{!{\vrule width .09em}l||*{6}{c|}c!{\vrule width .09em}}
 \toprule\midrule
 \multicolumn{8}{!{\vrule width .09em}c!{\vrule width .09em}}{Number of GW detections} \\
 \midrule\midrule
 \multirow{2}{*}{\textbf{Network}} & \multirow{2}{*}{$\Theta\leq\SI{20}{\degree}$} & \multirow{2}{*}{$\Theta\leq\SI{33}{\degree}$} & \multirow{2}{*}{$\Delta\psi\leq\SI{30}{\degree}$} & \multirow{2}{*}{$\Delta\iota\leq\SI{30}{\degree}$} & $\Theta\leq\SI{20}{\degree}$ \& & $\Theta\leq\SI{33}{\degree}$ \& & $\Theta\leq\SI{33}{\degree}$ \& \\
&  &  &  & & $\Delta\psi\leq\SI{30}{\degree}$ & $\Delta\psi\leq\SI{30}{\degree}$ & $\Delta\iota\leq\SI{30}{\degree}$\\
 \midrule
 LVKI O5 & 2 & 7 & 3 & 5 & 0 & 0 & 0\\
 LVKI Voyager & 74 & 148 & 100 & 78 & 8 & 26 & 1 \\
 ET & 1573 & 3774 & 1561 & 4007 & 26 & 63 & 54\\
 ET+2CE & 4680 & 12035 & 16973 & 21423 & 59 & 172 & 144\\
 \midrule\bottomrule
\end{tabular}
\caption{Summary of the results of GW detections relevant for the present work, at the various considered detector networks. For each configuration, in the second and third column we report the number of detections of events with $\Theta\leq\SI{20}{\degree}$ and $\Theta\leq\SI{33}{\degree}$, respectively. We use for convenience of notation the definition $\Theta = \min\{\iota, \SI{180}{\degree}-\iota\}$, thus e.g. $\Theta\leq\SI{20}{\degree}$ is equivalent to $\iota\leq\SI{20}{\degree}$ or $\iota\geq\SI{160}{\degree}$. The value $\Theta=\SI{33}{\degree}$ matches the one reconstructed for GW170817, which we take as a representative limit. In the fourth and fifth column we report the number of detected events with $1\sigma$ errors smaller than $\SI{30}{\degree}$ on $\psi$ and $\iota$, respectively. In the sixth and seventh column we report the number of events matching both the criteria $\Delta\psi\leq\SI{30}{\degree}$ and $\Theta\leq\SI{20}{\degree}$ and $\Theta\leq\SI{33}{\degree}$, respectively, while in the last column the events having both $\Theta\leq\SI{33}{\degree}$ and $\Delta\iota\leq\SI{30}{\degree}$ (no event is found with $\Theta\leq\SI{20}{\degree}$ and $\Delta\iota\leq\SI{30}{\degree}$, as can be seen from \autoref{fig:3G_prospects_iota}).}\label{tab:summary_GW}
\end{table}

\section{Conclusions and Discussion}\label{sec:conlcusions}

In this work we explored the scientific potential of combining information from gamma--ray polarization of short GRBs and detections of GWs from binary neutron star systems. We have shown that, although both measurements can make important contributions individually, a joint detection would significantly increase the scientific potential of polarimetry measurements. Whereas without GW measurements a large sample of GRB polarization measurements is required to constrain GRB emission models, a single joint detection would allow to exclude many if not all emission models. This result is illustrated in \autoref{tab:summary}. This table contains a summary of the predictions regarding the correlation between the polarization parameters of the prompt emission and the angles $\iota$ and $\psi$ defining the orientation of the binary system, measured with GW observations. For large off--axis viewing angles our results indicate that the emission model can be identified with a single GW measurement of $\iota$ and $\psi$ combined with a measurement of the polarization of the prompt emission. For small values of the inclination angle $\iota$, instead, a single measurement of the PD would allow to identify or discard synchrotron emission models with a toroidal magnetic field.

It is clear from the discussion in \autoref{sec:future_prospects} that detections with a joint measurement of both the PD, PA and the angles $\iota$ and $\psi$ are not likely to happen in the coming decade. Using instruments such as POLAR-2 and LEAP, highly constraining measurements in combination with GW data remain only possible in case of outlier events such as GW170817A. In the next decade, several measurements per year will become possible with instruments such as the ET and a third generation of polarimeters. For such events the relative error on $\iota$ and $\psi$ remains significant using GW data alone; however, this can be improved using EM follow--up measurements, as was the case for GW170817. Since the PA predicted by competing emission models can vary by as much as $\SI{90}{\degree}$, a precision of several 10's of degrees on the polarization angle $\psi$ would suffice to complement the polarization data. 

Finally, it should be noted that such joint observations could further benefit from high--resolution very long baseline interferometry (VLBI) measurements of the GRB afterglow, as the ones performed for GRB 170817A \citep{Mooley:2018qfh, Ghirlanda:2018uyx}. 
When $\Theta> 0$, the centroid of the afterglow radio surface brightness distribution moves in the plane of the sky, producing an apparently ‘super--luminal' motion.
For 170817A the $\Theta$ angle was measured to be $20\substack{+8 \\ -6}$ degrees \citep{Mooley:2018qfh}.
It could also be possible to directly measure $\psi$, at least for sources as close as GW170817, from the projected direction of this super--luminal motion away from its original detection location. Taking the measured centroid positions of the VLBI images of 170817A reported in \cite{Mooley:2018qfh} and \cite{Ghirlanda:2018uyx} the direction of the super--luminal motion can be fitted with a simple 2 parameter straight line. From such fits, using uniform priors for the angular parameters, an uncertainty on the $\psi$ angle of $7.0^\circ$ with a $68\%$ credibility (or $11.6^\circ$ with $90\%$ credibility) can be estimated.\footnote{Private communication with Om Salafia.} We can therefore conclude that VLBI measurements, at least for nearby events, can be highly complementary to gamma--ray polarization measurements. In addition, based on the precision found here, we encourage future measurements of the $\psi$ angle with VLBI and, when possible, their comparison to independent measurements of $\psi$ from GW data.

\begin{table}
\centering
\begin{tabular}{!{\vrule width .09em}p{3cm}||p{2.5cm}|p{3.5cm}|p{2cm}|p{2cm}!{\vrule width .09em}}
 \toprule\midrule
 \multicolumn{5}{!{\vrule width .09em}c!{\vrule width .09em}}{Emission model predictions} \\
 \midrule\midrule
 Model& PD ($\iota<\theta_c$) & PD ($\iota>\theta_c$) & PA ($\iota<\theta_c$) & PA ($\iota>\theta_c$)  \\
 \midrule
 Photosphere   & 0 & Low ($<20\%$) & -- & $\psi$\\
 Compton Drag   & 0 & Medium ($<40\%$) & -- & $\psi$\\
 Synch. Rand.   & 0 & Medium ($<40\%$) & -- & $\psi+\pi/2$\\
 Synch. Tor.   & High ($\sim50\%$)& 0 & $\psi+\pi/2$ & -- \\
 Synch. $B_\parallel$ & 0 & High ($>50\%$) & -- & random \\
 \midrule\bottomrule
\end{tabular}
\caption{\label{tab:summary} Qualitative predictions for the correlation between the PD and PA with the angles $\iota$ and $\psi$, for several emission models. The predictions are given for the case of a $\iota$ angle smaller and larger than the half opening angle of the jet ($\theta_c$). The PD values presented here are based on results presented in \cite{Gill2018}.}
\end{table}

\section*{Acknowledgements}

We thank Stefano Foffa, Michele Maggiore and Volodymyr Savchenko for many useful discussions. We also thank the coordinators of the ET OSB Division 4 for useful comments on the manuscript. We are indebted to Om Salafia for useful comments on the draft and for providing the estimate on the uncertainty on the polarization angle from VLBI measurements.
M.K. acknowledges the support of the Swiss National Science Foundation through the Ambizione program PZ00P2$\_$186119. F.I and M.M. are supported by  the  Swiss National Science Foundation, grant 200020$\_$191957, and  by the SwissMap National Center for Competence in Research. M.M. is supported by European Union's H2020 ERC Starting Grant No. 945155-GWmining and Cariplo Foundation Grant No. 2021-0555. Computations made use of the Yggdrasil cluster at the University of Geneva.

\bibliographystyle{aa}
\bibliography{bib}

\input{acronyms}

\newpage
\appendix
\section{Details on the Fisher matrix analysis of GW events}\label{sec:appendix_fisher}

The forecasts on the parameter estimation capabilities of GW detectors are obtained with a Fisher matrix analysis. This is used when a full parameter estimation is not possible, as in the case of large catalogs of events, and is valid in the linearized signal approximation, which is equivalent to the limit of large SNR (see \cite{Vallisneri:2007ev} for a detailed discussion). In this case, the likelihood $\mathcal{L}(s \,|\, \vb*{\theta})$ for data $s$ given source parameters $\vb*{\theta}$ can be approximated as a multivariate Gaussian distribution with covariance given by the inverse of the Fisher Information Matrix $\Gamma_{ij}$. For a detailed definition of this quantity, we refer to \cite{Iacovelli:2022bbs}, while we refer to \cite{Iacovelli:2022mbg} for a description of the package \texttt{GWFAST} used to compute the Fisher matrices in this paper.

\noindent The marginalized $1\sigma$ uncertainty for a parameter $\theta_i$ of the waveform is obtained in the Fisher approximation as
\begin{equation}
    \Delta \theta_i = \sqrt{\left[\Gamma^{-1}\right]_{ii} }
\end{equation}
As can be seen from the above definition, the calculation of the error involves the inversion of the Fisher matrix, so the accuracy in this operation is crucial for the reliability of the forecasts.
However, Fisher matrices with large dimensionality and with correlated parameters can show singularities, in the sense that they can have eigenvalues differing by many orders of magnitude. This matrices are ‘‘ill--conditioned'', in the sense that the condition number (i.e. the ratio of the largest to smallest eigenvalues) can be larger than the inverse of machine precision, in which case the inversion of the matrix is unreliable.
A commonly adopted procedure consists in discarding matrices with condition number larger than the inverse machine precision, or to check that the error on the computation of the inverse is under control (see the discussion in \cite{Iacovelli:2022bbs}). However, this procedure might introduce a bias when studying the distribution of the expected parameter estimation errors on a population of sources, since ill--conditioned matrices are associated to specific regions of the parameter space. In particular, events with values of the inclination angle close to $\SI{0}{\degree}$ or $\SI{180}{\degree}$ show systematically high condition numbers, since for those values the two polarizations of the GW signal become indistinguishable (see Fig. 1 of \cite{Iacovelli:2022bbs}). This is relevant for this study, since events with $\iota\sim0$ or $\iota\sim\pi$ are those interesting for the perspective of a joint detection with EM measurements.
In order to handle ill--conditioned matrices without discarding potentially significant events, we make the analysis more realistic by imposing a physically--motivated prior on the GW parameters. 
To do this, we proceed as follows. The posterior probability for source parameters $\vb*{\theta}$ given data $s$, in the linear signal/high SNR approximation, can be written explicitly as \citep{Iacovelli:2022bbs}

\begin{equation}\label{postGW}
    p(\vb*{\theta} \,|\, s) \propto \pi({\vb*{\theta}}) \times \mathcal{L}(s \,|\, \vb*{\theta})\, , \quad \mathcal{L}(s \,|\, \vb*{\theta}) \propto \exp \left[ -\frac{1}{2} \Gamma_{ij}  \big( \theta^{i}- \langle { \theta}^i \rangle\big) \big( \theta^{j}- \langle { \theta}^j \rangle\big) \right] \, ,
\end{equation}
where $\Gamma_{ij}$ is the Fisher matrix computed with \texttt{GWFAST}, $\langle {\theta}^i \rangle $ is the expected posterior mean on the $i$--th parameter -- which we take for simplicity to be coinciding with the true value -- and $\pi({\vb*{\theta}})$ is the prior. The prior choices on the waveform parameters are summarized in \autoref{tab:input_pars}.
Written in this form, the evaluation of the posterior does not require the explicit inversion of the Fisher Matrix, and one can extract samples from this distribution.
For each event in the simulated catalogues of detections, we draw samples from  \eqref{postGW} by sampling from the likelihood $\mathcal{L}(s \,|\, \vb*{\theta})$ defined in \eqref{postGW}, and rejecting samples outside the prior range. See Appendix C of \cite{Iacovelli:2022bbs} for more technical details.
Then, we quantify the $1\sigma$ marginal uncertainties on each parameter by computing the covariance of the samples.

\begin{table}[t]
    \centering
    \begin{tabular}{!{\vrule width .09em}c||c|c|c!{\vrule width .09em}}
    \toprule\midrule
        Parameter & Description &  Units & Prior range\\
        \midrule\midrule
         $m_1$ & detector--frame primary mass & \si{\Msun} & $(0,\, +\infty)$\\
         \midrule
         $m_2$ & detector--frame secondary mass   & \si{\Msun} & $(0,\, m_1]$\\
         \midrule
         $d_L$ & luminosity distance  & \si{\giga\parsec} & $(0,\, +\infty)$\\
         \midrule
         $\theta$, $\phi$  & polar and azimutal sky angles &rad & $[0,\, \pi]$, $[0,\, 2\pi]$ \\
         \midrule
         \multirow{2}{*}{$\iota$} & inclination angle w.r.t. orbital  & \multirow{2}{*}{rad} & \multirow{2}{*}{$[0,\, \pi]$}\\
        & angular momentum & &  \\
        \midrule
        $\psi$ & polarization angle  & rad & $[0,\, \pi]$ \\
         \midrule
        $t_c$ & time of coalescence GMST  & day &  $[0,\, 1]$\\
        \midrule
        $\Phi_c$ & phase at coalescence  & rad & $[0,\, 2\pi]$\\
        \midrule
        \multirow{2}{*}{$\chi_{i,z}$} & spin component of object $i=\{1,2\}$ & \multirow{2}{*}{--} & \multirow{2}{*}{$[-1,\,1]$} \\
        & along axis $z$ & &  \\
        \midrule
        \multirow{2}{*}{$\Lambda_{i}$} & adimensional tidal deformability  & \multirow{2}{*}{--} & \multirow{2}{*}{$[0,\, +\infty)$}\\
        & of object $i=\{1,2\}$ & &  \\
    \midrule\bottomrule
    \end{tabular}
    \caption{Summary of the parameters used to describe the GW signal of BNS in this paper, together with the prior range used when sampling from the posterior probability.}
    \label{tab:input_pars}
\end{table}

\end{document}

%% file: data_macros.tex
\def\addVAR#1#2{\expandafter\gdef\csname my@data@\detokenize{#1}\endcsname{#2}}
\def\VAR#1{%
  \ifcsname my@data@\detokenize{#1}\endcsname
    \csname my@data@\detokenize{#1}\expandafter\endcsname
  \else
    \expandafter\ERROR
  \fi
}

\addVAR{gw170817.SNR.LIGO_H|round(1)}{18.8}
\addVAR{gw170817.SNR.LIGO_L|round(1)}{26.4}
\addVAR{gw170817.SNR.Virgo|round(1)}{2.0}

%% file: acronyms.tex
\begin{acronym}
\acrodef{AG}[AG]{afterglow}
\acrodef{BAT}[BAT]{Burst Alert Telescope}
\acrodef{BB}[BB]{blackbody}
\acrodef{BBH}[BBH]{binary black hole}
\acrodef{BH}[BH]{black hole}
\acrodef{BNS}[BNS]{binary neutron star}
\acrodef{CR}[CR]{cosmic-ray}
\acrodef{EM}[EM]{electromagnetic}
\acrodef{EOS}[EOS]{equation of state}
\acrodef{FAP}[FAP]{false alarm probability}
\acrodef{FAR}[FAR]{false alarm rate}
\acrodef{GBM}[GBM]{Gamma-ray Burst Monitor}
\acrodef{GCN}[GCN]{Gamma-ray Coordinates Network}
\acrodef{GRB}[GRB]{gamma--ray burst}
\acrodef{GR}[GR]{General Relativity}
\acrodef{GW}[GW]{gravitational wave}
\acrodef{IBIS}[IBIS]{Imager on Board the INTEGRAL Satellite}
\acrodef{IBIS/Veto}[IBIS/Veto]{Veto shield of the Imager on Board the INTEGRAL Satellite}
\acrodef{INTEGRAL}[INTEGRAL]{INTErnational Gamma-Ray Astrophysics Laboratory}
\acrodef{IPN}[IPN]{InterPlanetary Network}
\acrodef{ISGRI}[ISGRI]{INTEGRAL Soft Gamma-Ray Imager}
\acrodef{JEM-X}[JEM-X]{Joint European X-Ray Monitor}
\acrodef{LIGO}[LIGO]{Laser Interferometer Gravitational-Wave Observatory}
\acrodef{MCMC}[MCMC]{Markov Chain Monte Carlo}
\acrodef{NS}[NS]{neutron star}
\acrodef{OMC}[OMC]{Optical Monitoring Camera}
\acrodef{PICsIT}[PICsIT]{PIxellated CsI Telescope}
\acrodef{RAVEN}[RAVEN]{Rapid VOEvent Coincidence Monitor}
\acrodef{SAA}[SAA]{South Atlantic Anomaly}
\acrodef{sGRB}[sGRB]{short gamma-ray burst}
\acrodef{SME}[SME]{Standard Model Extension}
\acrodef{SNR}[SNR]{signal--to--noise ratio}
\acrodef{SPI-ACS}[SPI-ACS]{SPectrometer onboard INTEGRAL - Anti-Coincidence Shield}
\acrodef{SPI}[SPI]{SPectrometer onboard INTEGRAL}
\end{acronym}